\documentstyle[12pt,epsfig]{article}
\catcode`\@=11
\@addtoreset{equation}{section}

\def\theequation{\thesection.\arabic{equation}}

\newtoks\@stequation

\def\subequations{\refstepcounter{equation}%
  \edef\@savedequation{\the\c@equation}%
  \@stequation=\expandafter{\theequation}
  \edef\@savedtheequation{\the\@stequation}
  \edef\oldtheequation{\theequation}%
  \setcounter{equation}{0}%
  \def\theequation{\oldtheequation\alph{equation}}}

\def\endsubequations{\setcounter{equation}{\@savedequation}%
  \@stequation=\expandafter{\@savedtheequation}%
  \edef\theequation{\the\@stequation}\global\@ignoretrue
  \vspace*{-12pt} \\}

\def\hybrid{\topmargin -20pt	\oddsidemargin 0pt
	\headheight 0pt	\headsep 0pt
	\textwidth 6.25in	
	\textheight 9.5in	
	\marginparwidth .875in
	\parskip 5pt plus 1pt	\jot = 1.5ex}

\def\baselinestretch{1.2}

\catcode`\@=11

\def\marginnote#1{}
\newcount\hour
\newcount\minute
\newtoks\amorpm
\hour=\time\divide\hour by60
\minute=\time{\multiply\hour by60 \global\advance\minute by-\hour}
\edef\standardtime{{\ifnum\hour<12 \global\amorpm={am}%
	\else\global\amorpm={pm}\advance\hour by-12 \fi
	\ifnum\hour=0 \hour=12 \fi
	\number\hour:\ifnum\minute<10 0\fi\number\minute\the\amorpm}}
\edef\militarytime{\number\hour:\ifnum\minute<10 0\fi\number\minute}
\def\draftlabel#1{{\@bsphack\if@filesw {\let\thepage\relax
   \xdef\@gtempa{\write\@auxout{\string
      \newlabel{#1}{{\@currentlabel}{\thepage}}}}}\@gtempa
   \if@nobreak \ifvmode\nobreak\fi\fi\fi\@esphack}
	\gdef\@eqnlabel{#1}}
\def\@eqnlabel{}
\def\@vacuum{}
\def\draftmarginnote#1{\marginpar{\raggedright\scriptsize\tt#1}}

\def\draft{\oddsidemargin -.2truein
	\def\@oddfoot{\sl preliminary draft \hfil
	\rm\thepage\hfil\sl\today\quad\militarytime}
	\let\@evenfoot\@oddfoot	\overfullrule 3pt
	\let\label=\draftlabel
	\let\marginnote=\draftmarginnote
   \def\@eqnnum{(\theequation)\rlap{\kern\marginparsep\tt\@eqnlabel}%
\global\let\@eqnlabel\@vacuum}  }

\def\preprint{\twocolumn\sloppy\flushbottom\parindent 2em
	\leftmargini 2em\leftmarginv .5em\leftmarginvi .5em
	\oddsidemargin -.5in	\evensidemargin -.5in
	\columnsep .4in	\footheight 0pt
	\textwidth 10.in	\topmargin  -.4in
	\headheight 12pt \topskip .4in
	\textheight 6.9in \footskip 0pt
	\def\@oddhead{\thepage\hfil\addtocounter{page}{1}\thepage}
	\let\@evenhead\@oddhead	\def\@oddfoot{}	\def\@evenfoot{} }

\def\titlepage{\@restonecolfalse\if@twocolumn\@restonecoltrue\onecolumn
     \else \newpage \fi \thispagestyle{empty}\c@page\z@
	\def\thefootnote{\fnsymbol{footnote}} }

\def\endtitlepage{\if@restonecol\twocolumn \else \newpage \fi
	\def\thefootnote{\arabic{footnote}}
	\setcounter{footnote}{0}}  

\catcode`@=12
\relax

\def\figcap{\section*{Figure Captions\markboth
	{FIGURECAPTIONS}{FIGURECAPTIONS}}\list
	{Figure \arabic{enumi}:\hfill}{\settowidth\labelwidth{Figure
999:}
	\leftmargin\labelwidth
	\advance\leftmargin\labelsep\usecounter{enumi}}}
 \relax
\def\tablecap{\section*{Table Captions\markboth
	{TABLECAPTIONS}{TABLECAPTIONS}}\list
	{Table \arabic{enumi}:\hfill}{\settowidth\labelwidth{Table
999:}
	\leftmargin\labelwidth
	\advance\leftmargin\labelsep\usecounter{enumi}}}
 \relax
\def\reflist{\section*{References\markboth
	{REFLIST}{REFLIST}}\list
	{[\arabic{enumi}]\hfill}{\settowidth\labelwidth{[999]}
	\leftmargin\labelwidth
	\advance\leftmargin\labelsep\usecounter{enumi}}}
 \relax
\def\Q{{\mathchoice
{\setbox0=\hbox{$\displaystyle\rm Q$}\hbox{\raise 0.15\ht0\hbox to0pt
{\kern0.4\wd0\vrule height0.8\ht0\hss}\box0}}
{\setbox0=\hbox{$\textstyle\rm Q$}\hbox{\raise 0.15\ht0\hbox to0pt
{\kern0.4\wd0\vrule height0.8\ht0\hss}\box0}}
{\setbox0=\hbox{$\scriptstyle\rm Q$}\hbox{\raise 0.15\ht0\hbox to0pt
{\kern0.4\wd0\vrule height0.7\ht0\hss}\box0}}
{\setbox0=\hbox{$\scriptscriptstyle\rm Q$}\hbox{\raise 0.15\ht0\hbox to0pt
{\kern0.4\wd0\vrule height0.7\ht0\hss}\box0}}}}
\def\C{{\mathchoice
{\setbox0=\hbox{$\displaystyle\rm C$}\hbox{\hbox to0pt
{\kern0.4\wd0\vrule height0.9\ht0\hss}\box0}}
{\setbox0=\hbox{$\textstyle\rm C$}\hbox{\hbox to0pt
{\kern0.4\wd0\vrule height0.9\ht0\hss}\box0}}
{\setbox0=\hbox{$\scriptstyle\rm C$}\hbox{\hbox to0pt
{\kern0.4\wd0\vrule height0.9\ht0\hss}\box0}}
{\setbox0=\hbox{$\scriptscriptstyle\rm C$}\hbox{\hbox to0pt
{\kern0.4\wd0\vrule height0.9\ht0\hss}\box0}}}}

\font\fivesans=cmss10 at 4.61pt
\font\sevensans=cmss10 at 6.81pt
\font\tensans=cmss10
\newfam\sansfam
\textfont\sansfam=\tensans\scriptfont\sansfam=\sevensans\scriptscriptfont
\sansfam=\fivesans
\def\sans{\fam\sansfam\tensans}
\def\Z{{\mathchoice
{\hbox{$\sans\textstyle Z\kern-0.4em Z$}}
{\hbox{$\sans\textstyle Z\kern-0.4em Z$}}
{\hbox{$\sans\scriptstyle Z\kern-0.3em Z$}}
{\hbox{$\sans\scriptscriptstyle Z\kern-0.2em Z$}}}}

\mathchardef\endbar="375

\makeatletter
\newcounter{pubctr}
\def\publist{\@ifnextchar[{\@publist}{\@@publist}}
\def\@publist[#1]{\list
	{[\arabic{pubctr}]\hfill}{\settowidth\labelwidth{[999]}
	\leftmargin\labelwidth
	\advance\leftmargin\labelsep
	\@nmbrlisttrue\def\@listctr{pubctr}
	\setcounter{pubctr}{#1}\addtocounter{pubctr}{-1}}}
\def\@@publist{\list
	{[\arabic{pubctr}]\hfill}{\settowidth\labelwidth{[999]}
	\leftmargin\labelwidth
	\advance\leftmargin\labelsep
	\@nmbrlisttrue\def\@listctr{pubctr}}}
 \relax
\makeatother
\newskip\humongous \humongous=0pt plus 1000pt minus 1000pt
\def\caja{\mathsurround=0pt}
\def\eqalign#1{\,\vcenter{\openup1\jot \caja
	\ialign{\strut \hfil$\displaystyle{##}$&$
	\displaystyle{{}##}$\hfil\crcr#1\crcr}}\,}
\newif\ifdtup

\relax
\hybrid
\def\a{\alpha}
\def\b{\beta}

\def\e{\epsilon}
\def\m{\mu}

\def\t{\tau}

\def\s{\sigma}
\def\l{\lambda}
\def\o{\omega}
\def\ho{\hat{\omega}}
\def\hop{\hat{\omega}^+}
\def\hom{\hat{\omega}^-}

\def\bs{\bar{s}}

\def\tL{\tilde{L}}

\def\tc{\tilde{c}}
\def\tP{\tilde{\Phi}}
\def\tT{\tilde{T}}

\def\J{{\cal J}}

\def\P{{\cal P}}

\def\C{{\cal C}_{g/h}}
\def\cO{{\cal O}}

\def\pa{\partial}
\def\dif{\partial}
\def\difb{\bar{\partial}}
\def\dbar{\bar{\partial}}
\def\del{\nabla}
\def\de{\nabla}
\def\delh{\hat{\nabla}}

\def\delhp{\hat{\nabla}^+}
\def\delhm{\hat{\nabla}^-}
\def\delb{\bar{\nabla}}
\def\deb{\bar{\nabla}}

\def\xx{\hbox{ }^*_*}

\def\ok{\cO (k^{-2} )}

\def\thefootnote{\fnsymbol{footnote}}
\def\be{\begin{equation}}
\def\ee{\end{equation}}
\def\bs{\begin{subequations}}
\def\es{\end{subequations}}
\def\ben{\begin{enumerate}}
\def\een{\end{enumerate}}

\def\vs{\vskip}

\def\ed{\end{document}}

\def\bibtem#1{\bibitem{#1} }

\def\tr{{\rm Tr}}
\def\oa{ {\cal O}(\alpha ')}
\def\oaa{ {\cal O}(\alpha'^2)}
\def\P{\Pi}
\def\BP{\bar{\Pi}}
\def\Pb{\bar{\Pi}}
\def\half{\frac{1}{2}}
\def\zb{\bar{z}}
\def\wb{\bar{w}}
\def\kb{\bar{k}}
\def\pb{\bar{p}}
\def\fbb{\Pb^c \Pb^d \frac{(\zb-\wb)^2}{(z-w)^4} }
\def\frb{\P^c \Pb^d \frac{(\zb-\wb)}{(z-w)^3} }
\def\frr{\P^c \P^d \frac{1}{(z-w)^2} }
\def\frrg{\P^c \P^d \frac{G(z,w)}{(z-w)^2} }
\def\frrrg#1{\P^c \P^d \frac{G(z,w) #1}{(z-w)^2} }
\def\frrd{\P^c \P^d \frac{G(0,0)}{(z-w)^2} }
\def\bea{\begin{eqnarray}}
\def\ena{\end{eqnarray}}  \def\eea{\end{eqnarray}}
\def\nonu{\nonumber \\}

\begin{document}
\begin{titlepage}
\begin{center}

\hfill CERN-TH/96-132 \\
\hfill UCB-PTH-96/20 \\ 
\hfill LBNL-38860 \\  
\hfill ITP-SB-96-25 \\
\hfill hep-th/9606025  \\

\vskip .15in

{\large \bf  
Unified Einstein-Virasoro Master Equation\\
in the General Non-Linear Sigma Model}
\vskip .125in
{\bf J. de Boer}
\footnote{e-mail address: deboer@insti.physics.sunysb.edu}
\vskip .1in
{\em Institute for Theoretical Physics, SUNY
at Stony Brook\\
Stony Brook, NY 11794-3840, USA}
\\
\vskip .15in
and
\vskip .125in
{\bf M.B. Halpern}
\footnote{On leave from the Department of Physics, University of California,
 Berkeley, CA 94720, USA.} 
\footnote{e-mail address: mbhalpern@theor3.lbl.gov, 
halpern@vxcern.cern.ch}
\\
\vskip .1in
{\em Theory Division, CERN, CH-1211,
Geneva 23, SWITZERLAND} 
\end{center}

\vskip .17in

\begin{center} {\bf ABSTRACT } \end{center}


\noindent
{\small 
The Virasoro master equation (VME) describes the general affine-Virasoro
construction $T=L^{ab}J_aJ_b+iD^a \dif J_a$ in the operator algebra of the WZW
model, where $L^{ab}$ is the inverse inertia tensor and $D^a $ is
the improvement vector. In this paper, we generalize this construction to
find the general (one-loop) Virasoro construction in the operator
algebra of the general non-linear sigma model. The result is a unified 
Einstein-Virasoro master equation which couples the spacetime spin-two
field $L^{ab}$ to the background fields of the
sigma model.
For a particular solution $L_G^{ab}$, the unified system reduces to the 
canonical stress tensors and 
conventional Einstein equations of the sigma model, and the
system reduces to the 
general affine-Virasoro construction and the
VME when the sigma model is taken to be the WZW
action. 
More generally, the unified system describes a space of conformal
field theories which is presumably much larger than the sum 
of the general affine-Virasoro construction and the 
sigma model with its canonical stress tensors. 
We also 
discuss a number of algebraic and geometrical properties of the system,
including its relation to an unsolved problem in the theory of $G$-structures
on manifolds with torsion.}
 \vskip 0.3cm
\noindent
CERN-TH/96-132\\
\noindent
April 1996 
 \end{titlepage}
\vfill
\eject
\def\baselinestretch{1.2}
\baselineskip 14 pt
\noindent
\setcounter{equation}{0}
\tableofcontents
\def\baselinestretch{1.2}
\baselineskip 16 pt
\section{Introduction}
There have been two successful approaches to the construction of conformal
field theories,
\begin{itemize}
\item The general affine-Virasoro construction [1--7]
\item The general non-linear sigma model [8--13]
    \hfill (1.1)
\end{itemize}
\addtocounter{equation}{1}
but, although both approaches have been formulated as Einstein-like
systems \cite{sig9,gme}, the
relation between the two has remained unclear.

In the general affine-Virasoro construction, a large class of exact 
Virasoro operators \cite{hk,rus}
\bs\be
              T= L^{ab} \xx J_a J_b \xx + iD^a \dif J_a
\ee\be
a,b=1 \ldots \dim(g)
\ee
\label{avc} \es
are constructed as quadratic forms in the currents $J$ of the general
affine Lie algebra \cite{km,bh}. The coefficients $L^{ab}=L^{ba}$ and $D^a$ are
called the inverse inertia tensor and the improvement vector
respectively. The general construction is summarized \cite{hk,rus}
by the (improved)
Virasoro master equation (VME) for $L$ and $D$, and this approach is
the basis of irrational conformal field theory \cite{rev} which includes the
affine-Sugawara [15--18]
  and coset constructions \cite{bh,h1,gko} as a small subspace. The
construction (\ref{avc}) can also be considered as the 
general Virasoro construction in the operator algebra of the 
WZW model \cite{nov,wit},
which is the field-theoretic realization of the affine algebras.
See Ref \cite{rev} for a more detailed
history of affine Lie algebra and the affine-Virasoro
constructions.

In the general non-linear sigma model, a large class of Virasoro operators
\cite{sig12}
\bs
\be
T=-\frac{1}{2\a '} G_{ij} \dif x^i \dif x^j + {\cal O}(\a '^0) =
-\frac{1}{2\a '} G^{ab} \Pi_a \Pi_b + {\cal O}(\a '^0) 
\ee
\be
G^{ab}=e_i{}^a G^{ij} e_j{}^b , \qquad \Pi_a = G_{ab} e_i{}^b \dif x^i
\ee
\be
i,j,a,b=1,\ldots, \dim(M)
\ee \label{jjj2} \es
is constructed in a semiclassical expansion on an arbitrary manifold $M$,
where $G_{ij}$ is the metric on $M$ and $G^{ab}$ is the inverse of 
the tangent space metric. 
These are the canonical or conventional
stress tensors of the sigma model and this construction is summarized 
\cite{sig9,sig12} by the Einstein
equations of the sigma model, which couple the metric $G$, the
antisymmetric tensor field $B$ and the dilaton $\Phi$. In what
follows we refer to these equations as the conventional Einstein equations
of the sigma model, to distinguish them from the generalized 
Einstein equations obtained below.

In this paper, we unify these two approaches, using the fact that
the WZW action is a special case of the general sigma model. More
precisely, we study the general Virasoro construction
\bs\be
T=-\frac{1}{\a '} L_{ij} \dif x^i \dif x^j + {\cal O}(\a '^0) =
-\frac{1}{\a '} L^{ab} \Pi_a \Pi_b + {\cal O}(\a '^0) 
\ee\be
i,j,a,b = 1 , \ldots , \dim(M)
\ee \label{jjj3} \es
at one loop in the operator algebra of the general sigma model, where 
$L$ is a symmetric second-rank spacetime tensor field,
the inverse inertia tensor, which is to be determined.
The unified construction is described by a system of equations which we
call
\begin{itemize}
\item the Einstein-Virasoro master equation
\end{itemize}
of the general sigma model. This system, which is summarized 
in Section~4, describes
the covariant coupling of the spacetime fields
$L$, $G$, $B$ and $\Phi_a$, where the vector field $\Phi_a$ generalizes
the derivative $\del_a \Phi$ of the dilaton $\Phi$.

The unified system contains as special cases the two constructions in
(1.1): For the particular solution
\be L^{ab}=L_G^{ab} = \frac{G^{ab}}{2} + \oa ,
\qquad \Phi_a = \Phi_a^G = \del_a \Phi   \ee
the general stress tensors (\ref{jjj3}) reduce to the conventional 
stress tensors (\ref{jjj2}) and the Einstein-Virasoro master equation
reduces to the conventional Einstein equations of the sigma model.
Moreover, the unified system reduces to the general affine-Virasoro
construction and the VME when the sigma model is taken to be the WZW
action. In this case we find that the contribution of $\Phi_a$ to
the unified system is precisely the known improvement term of the
VME. 

More generally, the unified system describes a space of conformal
field theories which is presumably much larger than the sum of the
general affine-Virasoro construction and the sigma model with its 
canonical stress tensors.

The system exhibits a number of interesting algebraic and geometric
properties, including $K$-conjugation covariance \cite{bh,h1,gko,hk}
 and an important
connection with the theory of $G$-structures \cite{hol12} on manifolds with torsion.
By comparison with the known exact form of the VME, we are also able to
point to a number of candidates for exact relations in the general
sigma model.

\section{Background}

To settle notation and fix concepts which will be important below, we begin
with a brief review of the two known constructions,
\begin{itemize}
\item The general affine-Virasoro construction 
\item The general non-linear sigma model.
\end{itemize}

\vfill
\eject

\subsection{The general Affine-Virasoro Construction}

\vs .4cm
\noindent \underline{The improved VME}
\vs .3cm 

The general affine-Virasoro construction begins with the currents of a general
affine Lie algebra \cite{km,bh}
\be
J_a(z) J_b(w) = \frac{G_{ab}}{(z-w)^2} + 
\frac{i f_{ab}{}^c J_c(w) }{ z-w} + {\rm reg.}
\label{eq1}
\ee
where $a,b=1\ldots \dim g$ and $f_{ab}{}^c$ are the structure constants
of $g$. For simple $g$, the central term in (\ref{eq1}) has the
form $G_{ab}=k\eta_{ab}$ where $\eta_{ab}$ is the Killing metric of
$g$ and $k$ is the level of the affine algebra. Then the general 
affine-Virasoro construction is \cite{hk}
\be
              T= L^{ab} \xx J_a J_b \xx + iD^a \dif J_a
\label{avc2} \ee
where the coefficients
$L^{ab}=L^{ba}$ and $D^a$ are the inverse inertia tensor and the 
improvement vector respectively. The stress tensor $T$ is a Virasoro
operator
\be
T(z) T(w) = \frac{c/2}{(z-w)^4} + \frac{2 T(w)}{(z-w)^2} + 
\frac{\dif_w T(w)}{z-w} + {\rm reg.} 
\label{ttope}
\ee
iff the improved VME \cite{hk}
\bs
\be
L^{ab} = 2 L^{ac} G_{cd} L^{db} - L^{cd} L^{ef} f_{ce}{}^a f_{df}{}^b
-L^{cd} f_{ce}{}^f f_{df}{}^{(a} L^{b)e} - f_{cd}{}^{(a} L^{b)c} D^d
\label{VME1}
\ee
\be
D^a (2 G_{ab} L^{be} + f_{ab}{}^d L^{bc} f_{cd}{}^e ) = D^e
\label{VME2}
\ee
\be c=2G_{ab} (L^{ab} + 6 D^a D^b) \label{VME3} \ee
\label{VMEall}
\es
is satisfied\footnote{Our convention is $A^{(a} B^{b)} = A^a B^b + A^b B^a$,
 $A^{[a} B^{b]} = A^a B^b - A^b B^a$.} by $L$ and $D$, and
the central charge of the 
construction is given in (\ref{VME3}). The unimproved VME \cite{hk,rus} is 
obtained by setting the improvement vector $D$ to zero. 

The improved VME has been identified \cite{gme} as an Einstein-Maxwell system with
torsion on the group manifold, with left-invariant inverse metric
\bs
\be
G^{ij} = e_a{}^i L^{ab} e_b{}^j \label{ein1}
\ee
\be D_i L^{ab} = 0 \label{ein2}
\ee
\label{iii}
\es
where $e_a{}^i$ is the inverse left-invariant vielbein and $L^{ab}$
is the inverse of the tangent space metric.

Another property that will be useful below is that the improved
VME is covariant under automorphisms of $g$
\bs 
\be 
(L')^{ab}=L^{cd} \rho_c{}^a \rho_d{}^b, \qquad (D')^a=D^b \rho_b{}^a
\ee
\be
c(L',D')=c(L,D),\qquad \rho\in{\rm Aut}(g)
\ee
\be \rho_a{}^c \rho_b{}^d G_{cd} = G_{ab}, \qquad 
\rho_a{}^c \rho_b{}^d f_{cd}{}^e = f_{ab}{}^c \Omega_c{}^e
\ee
\es
with invariant central charge.

\vs .4cm
\noindent \underline{$K$-conjugation invariance}
\vs .3cm 

A central property of the VME at zero improvement\footnote{$K$-conjugation
in the presence of the improvement term is discussed in Section~5.9}
is $K$-conjugation covariance \cite{bh,h1,gko,hk}
which says that all solutions come in
$K$-conjugate pairs $L$ and $\tilde{L}$,
\bs
\be
L^{ab}+\tilde{L}^{ab} = L_g^{ab} .
\ee
\be
T+\tilde{T} = T_g
\ee\be
c+\tilde{c} = c_g 
\ee\be
T(z)\tilde{T}(w) = {\rm reg.}
\ee
\es
whose $K$-conjugate stress tensors $T,\tilde{T}$ commute and add
to the affine-Sugawara construction [15--18] on $g$
\be
T_g = L_g^{ab} \xx J_a J_b \xx .
\label{affsug}
\ee
For simple $g$, the inverse inertia tensor and the central charge of the
affine-Sugawara construction are
\bs
\be
L_g^{ab} = \frac{\eta^{ab}}{2k+Q_g} = \frac{\eta^{ab}}{2k} + \ok = 
\frac{G^{ab}}{2} + \ok 
\label{affa}
\ee\be
c_g = \frac{\dim(g)}{1+\frac{Q_g}{2k} } = \dim(g)
\left( 1 - \frac{Q_g}{2k} \right) + \ok
\label{affb}
\ee\be
Q_g \eta_{ab} = -f_{ac}{}^d f_{bd}{}^c
\label{affc}
\ee
\label{afff}
\es
where $\eta^{ab}$ is the inverse Killing metric of $g$ and $Q_g$ is the 
quadratic Casimir of the adjoint. $K$-conjugation covariance can be used
to generate new solutions $\tilde{L}=L_g-L$ from old solutions $L$ and
the simplest application of the covariance generates the coset constructions
\cite{bh,h1,gko} as $\tilde{L}=L_g-L_h=L_{g/h}$.

\vs .4cm
\noindent \underline{Semiclassical expansion}
\vs .3cm 

At zero improvement, the high-level or semiclassical expansion \cite{hl,rev}
 of
the VME has been studied in some detail. On simple $g$, the leading term
in the expansion has the form
\bs\be
L^{ab} = \frac{P^{ab}}{2k} + \ok, \qquad c = {\rm rank}(P) + 
{\cal O}(k^{-1}) 
\ee\be
P^{ac} \eta_{cd} P^{db} = P^{ab}
\ee
\label{expan1}
\es
where $P$ is the high-level projector of the $L$ theory. These are the 
solutions of the classical limit of the VME,
\be L^{ab} = 2 L^{ac} G_{cd} L^{db} + \ok 
\label{expan2}
\ee
but a semiclassical quantization condition \cite{hl} provides a 
restriction on the allowed projectors. In the partial classification
of the space of solutions by graph theory \cite{gt,ggt,rev}, the projectors 
$P$ are closely related to the
adjacency matrices of the graphs.

\vs .4cm
\noindent \underline{Irrational conformal field theory}
\vs .3cm 

Given also a set of antiholomorphic currents $\bar{J}_a$, $a=1\ldots
\dim(g)$, there is a corresponding antiholomorphic Virasoro construction
\be
\bar{T}= L^{ab} \xx \bar{J}_a \bar{J}_b \xx + iD^a \dif \bar{J}_a
\label{avc3} \ee
with $\bar{c}=c$. Each pair of stress tensors $T$ and $\bar{T}$ then
defines a conformal field theory (CFT) labelled by $L$ and $D$. 
Starting from the modules of affine $g \times g$, the 
Hilbert space
of a particular CFT is obtained \cite{hy,rp,rev} by modding out by the
local symmetry of the Hamiltonian.

It is known that the CFTs of the master equation have generically
irrational central charge, even when attention is restricted to the
space of unitary theories, and the study of all the CFTs of the 
master equation is called irrational conformal field theory (ICFT),
which contains the affine-Sugawara and coset constructions as a 
small subspace. 

In ICFT at zero improvement, 
world-sheet actions are known for the following cases: the 
affine-Sugawara constructions (WZW models \cite{nov,wit}),
the coset constructions (spin-one gauged WZW models \cite{br}) and the
generic ICFT (spin-two gauged WZW models \cite{hy,ts3,bch}). The spin-two
gauge symmetry of the generic ICFT is a consequence of
$K$-conjugation covariance.

See Ref \cite{rev} for a comprehensive review of ICFT, 
and Ref \cite{blocks} for a 
recent construction of a set of
semiclassical blocks and correlators in ICFT.

\vs .4cm
\noindent \underline{WZW model}
\vs .3cm 

The stress tensors (\ref{avc2}) and (\ref{avc3}) of the general
affine-Virasoro construction are general quadratic forms in the currents
of affine $g \times g$. As such, these stress tensors may also be
considered as the general Virasoro construction in the operator
algebra of the WZW model on the group manifold $G$, where $g$ is
the algebra of $G$.


For simple $G$, the Minkowski space WZW action is \cite{nov,wit}
\bs\be
S_{WZW} = -\frac{k}{8\pi\chi} \int dt d\sigma \tr(g^{-1}
\dif_+ g g^{-1} \dif_- g ) - \frac{k}{12 \pi \chi}
\int \tr (g^{-1} d g)^3
\label{wzwact}
\ee
\be
g(T)\in G, \qquad \tr(T_aT_b) = \chi \eta_{ab}
\ee
\be \partial_{\pm} = \partial_t \pm \partial_{\sigma} 
\ee
\es
where $g(T)$ is the group element in irrep $T$ of $g$, $\eta_{ab}$ is the
Killing metric of $g$ and $k$ is the level of the affine algebra.
The group element $g$ is dimensionless, and we may introduce dimensionless
coordinates $y^i$, $i=1\ldots \dim(g)$, such that
\bs\be
-i g^{-1} \frac{\dif}{\dif y^i} g = e_i{}^a T_a, \qquad
-i g \frac{\dif}{\dif y^i} g^{-1} = \bar{e}_i{}^a T_a 
\label{invvielbein}
\ee\be
g T_a g^{-1} = \Omega_a{}^b T_b, \qquad \bar{e}_i{}^a = - e_i{}^b
 \Omega_b{}^a
\label{aux2}
\ee\be
\Omega_a{}^c \Omega_d{}^b G_{cd} = G_{ab} , \qquad
\Omega_a{}^c \Omega_b{}^d f_{cd}{}^e = f_{ab}{}^c \Omega_c{}^e 
\label{aux3}
\ee\be
\frac{\dif}{\dif y^i} \Omega_a{}^b + e_i{}^d f_{da}{}^c \Omega_c{}^b = 0
\label{aux4}
\ee
\label{newcoords}
\es
where $e$ and $\bar{e}$ are 
respectively
the left and right invariant vielbein on $G$
and $\Omega$ is the adjoint action of $g$.

Continuing in this direction, one rewrites the WZW action as a special case
of the general non-linear sigma model, a subject to which we will return in
the following section.

\subsection{The general non-linear sigma model}

The general non-linear sigma model has been extensively studied in 
\cite{sig1,sig1a,sig2,sig3,sig4,sig5,sig6,sig7,sig8,sig9,sig11,sig10,sig12,sig13}.

The Euclidean action of the general non-linear sigma model is
\bs\be
S=\frac{1}{2\a '} \int d^2 z  (G_{ij} + B_{ij}) \dif x^i \dbar x^j
\label{sigmaact}
\ee\be
d^2z=\frac{dxdy}{\pi}, \qquad z=x+iy
\label{measure}
\ee\be
H_{ijk} = \dif_i B_{jk} + \dif_j B_{ki} + \dif_k B_{ij}.
\ee\es
Here $x^i$, $i=1\ldots \dim(M)$ are coordinates with the dimension
of length on a general manifold $M$ and $\a'$, with dimension length
squared, is the string tension or Regge slope. 
The fields $G_{ij}$ and $B_{ij}$ are 
the (covariantly constant) metric and antisymmetric tensor field on $M$.

We also introduce a covariantly constant vielbein $e_i{}^a$, $a=1\ldots
\dim(M)$ on $M$ and use it to translate between Einstein and 
tangent-space indices, e.g.
\be G_{ij}=e_i{}^a G_{ab} e_j{}^b \ee
where $G_{ab}$ is the covariantly constant metric on tangent space. Covariant
derivatives are defined as
\bs\be
\del_i v_a = \dif_i v_a - \omega_{ia}{}^b v_b , \qquad
\del_i v^a = \dif_i v^a + v^b \omega_{ib}{}^a
\ee\be
\del_i G_{ab} = \dif_i G_{ab} = \nabla_i e_j{}^a =
0 
\ee\be
[\del_i,\del_j] v^a = R_{ijb}{}^a v^b
\ee\be
R_{ija}{}^b= (\dif_i \o_j - \dif_j \o_i - [\o_i,\o_j])_a{}^b
\ee\be
\del_a v_b = e_a{}^i \del_i v_b = \dif_a v_b - \o_{ab}{}^c v_c
\ee\es
where $\o$ is the spin connection, $R_{ija}{}^b$ is the
Riemann tensor and $R_{ab}=R_{acb}{}^c$ is the Ricci tensor.
It will also be convenient to define the generalized connections
and covariant derivatives with torsion,
\bs\be
\delh^{\pm}_i v_a = \dif_i v_a - \hat{\o}^{\pm}_{ia}{}^b v_b
\label{tordefa}
\ee\be
\ho^{\pm}_{ia}{}^b=\o_{ia}{}^b \pm \frac{1}{2} H_{ia}{}^b
\label{tordefb}
\ee\be
\delh^{\pm}_i G_{ab} = \delh^{\pm}_j e_i{}^a = 0
\label{tordefc}
\ee\be
\hat{R}^{\pm}_{ija}{}^b= (\dif_i \ho^{\pm}_j-\dif_j \ho^{\pm}_i
-[\ho_i^{\pm},\ho_j^{\pm}])_a{}^b
\label{tordefd}
\ee\be
[\delh_a^{\pm},\delh_b^{\pm}]v^c=\mp H_{ab}{}^d \delh_d^{\pm} v^c
+\hat{R}^{\pm}_{abd}{}^c v^d
\label{tordefe}
\ee\be
R^{\pm}_{cdab} = R_{cdab} \pm \frac{1}{2} (\del_c H_{dab} -
\del_d H_{cab} ) + \frac{1}{4} (H_{ca}{}^f H_{fdb} -
H_{da}{}^f H_{fcb} ) 
\label{tordeff}
\ee\be
\hat{R}^{\pm} = \hat{R}^{\pm}_a{}^a = R+\frac{1}{4} H^2, 
\qquad (H^2)_{ij} = H_i{}^{kl} H_{jkl}
\ee
\label{tordef}
\es
where $\ho^{\pm}_{iab}$ is antisymmetric under $(a,b)$ interchange and
$\hat{R}^{\pm}_{ijab}$ is pairwise antisymmetric in $(i,j)$ and $(a,b)$.

Following Banks, Nemeschansky and Sen \cite{sig12}, the 
canonical or
conventional stress tensors
of the general sigma model have the form
\bs\be
T_G = -\frac{G_{ij}}{2\a '} \dif x^i \dif x^j + \dif^2 \Phi + T_1
+ {\cal O}(\a ')
\ee\be
\qquad  = -\frac{G^{ab}}{2\a '} \Pi_a \Pi_b + \dif^2 \Phi + T_1
+ {\cal O}(\a ')
\ee\be
\bar{T}_G = -\frac{G_{ij}}{2\a '} \dbar x^i \dbar x^j + \dbar^2 \Phi 
+ \bar{T}_1 + {\cal O}(\a ')
\ee\be
\qquad  = -\frac{G^{ab}}{2\a '} \bar{\Pi}_a \bar{\Pi}_b + \dbar^2 \Phi + 
\bar{T}_1
+ {\cal O}(\a ')
\ee\be
\Pi_a=G_{ab} e_i{}^b \dif x^i , \qquad \bar{\Pi}_a = G_{ab} e_i{}^b 
\dbar x^i
\label{defpi}
\ee \label{jjj19} \es
where $\Phi$ is the dilaton and $T_1,\bar{T}_1$ are finite one-loop 
counterterms which depend on the renormalization scheme.
The condition that $T_G$ and $\bar{T}_G$ are one-loop conformal
is \cite{sig9}
\bs\be
R_{ij}+\frac{1}{4} (H^2)_{ij} - 2 \del_i \del_j \Phi = 
{\cal O}(\a ') 
\label{conf1}
\ee\be
\del^k H_{kij} - 2 \del^k \Phi H_{kij} = {\cal O}(\a')
\label{conf2}
\ee\be
c_G=\bar{c}_G = \dim(M) + 3 \alpha' 
(4 |\del \Phi|^2 - 4 \del^2 \Phi + R + \frac{1}{12} H^2) + 
{\cal O}(\a '^2)
\label{conf3}
\ee
\label{conf}
\es
where (\ref{conf1}) and (\ref{conf2}) are the conventional
Einstein equations of 
the sigma model and (\ref{conf3}) is the central charge of the
construction. The result for the central charge includes two-loop 
information, but covariant constancy of the field-dependent part
of the central charge follows by Bianchi identities from the
Einstein equations, so all three relations 
in (\ref{conf})
can be obtained with a little
thought from the one-loop calculation. It will also be useful to note that
the conventional
Einstein equations (\ref{conf1}),(\ref{conf2}) can be written in
either of two equivalent forms
\be
\hat{R}^{\pm}_{ij} - 2 \delh^{\pm}_i \delh^{\pm}_j \Phi = 
{\cal O}(\a ')
\label{conf4}
\ee
by using the generalized quantities (\ref{tordef}) with torsion.

\vs .4cm
\noindent \underline{WZW data}
\vs .3cm 

The WZW action (\ref{wzwact}) is a special case of the general sigma
model (\ref{sigmaact}) on a group manifold $G$. Identifying the
vielbein $e$ on $M$ with the left-invariant vielbein $e$
on $G$ in (\ref{invvielbein}), we find for WZW that
\be
x^i=\sqrt{\a'} y^i, \qquad G_{ab} = k \eta_{ab}, \qquad
H_{ab}{}^c = \frac{1}{\sqrt{\a'}} f_{ab}{}^c
\label{wzwdata}
\ee
where $y^i$, $i=1\ldots \dim(g)$ are the dimensionless 
coordinates introduced for WZW in (\ref{newcoords}), 
$f_{ab}{}^c$ and $\eta_{ab}$ are the structure constants and the Killing
metric of $g$ and $k$ is the level of the affine algebra. From this data,
one computes
\bs\be
\o_{ab}{}^c = -\frac{1}{2\sqrt{\a'}} f_{ab}{}^c 
\ee\be
R_{abc}{}^d = -\frac{1}{4\a'}f_{ab}{}^e f_{ec}{}^d,
\qquad R_{ab} = -\frac{Q_g}{4\a'} \eta_{ab}, \qquad
R=-\frac{Q_g}{4\a' k} \dim(g)
\ee\be
(H^2)_{ab} = \frac{Q_g}{\a'} \eta_{ab}, \qquad H^2=\frac{k Q_g \dim(g)}{\a'}
\ee\es
where $Q_g$ is the quadratic Casimir of the adjoint. Finally, one finds
\bs\be
\ho^+_{ab}{}^c = 0, \qquad \ho^-_{ab}{}^c = -\frac{1}{\sqrt{\a'}} f_{ab}{}^c
 \label{auxn17} \ee
\be
\hat{R}^{\pm}_{ija}{}^b = 0
\label{wzwrhat}
\ee\es
for the generalized quantities (\ref{tordef}) with torsion. Manifolds with 
vanishing generalized Riemann tensors are called parallelizable \cite{sig6,sig11}.

The relations between the classical WZW currents $J$,$\bar{J}$ and
the quantities $\P$,$\BP$ in (\ref{defpi}) are
\bs\be
J_a = ik \eta_{ab} e_i{}^b \dif y^i = \frac{i}{\sqrt{\a'}} \Pi_a
\label{auxn1a} \ee\be
\bar{J}_a = ik \eta_{ab} \bar{e}_i{}^b \dbar y^i = \frac{-i}{\sqrt{\a'}} 
(\Omega^{-1})_a{}^b \BP_b
\label{auxn1b} \ee \label{auxn1} \es
where $e$ and $\bar{e}$ are the left and right invariant vielbeins in 
(\ref{invvielbein}) and $\Omega$ is the adjoint action of $g$. It follows
that the classical limits of the affine-Sugawara stress tensors
\bs\be
T_G \rightarrow T_g = -\frac{L_g^{ab}}{\a'} \Pi_a \Pi_b + 
{\cal O}(\a'^0) = L_g^{ab} J_a J_b + {\cal O}(\a'^0) 
\label{auxn2a} \ee \be
\bar{T}_G \rightarrow \bar{T}_g = 
-\frac{L_g^{ab}}{\a'} \bar{\Pi}_a \bar{\Pi}_b + 
{\cal O}(\a'^0) = L_g^{ab} \bar{J}_a \bar{J}_b + {\cal O}(\a'^0) 
\label{auxn2b} \ee \be
L_g^{ab}  \approx \frac{G^{ab}}{2} = 
\frac{\eta^{ab}}{2k} 
\label{auxn2c} \ee \label{auxn2} \es
are correctly obtained as a special case of the conventional 
stress tensors (\ref{jjj19}). The result in (\ref{auxn2b}) follows because
the adjoint action is an inner automorphism of $g$.

We can also check at the WZW level that the affine-Sugawara stress tensors
(\ref{auxn2})
are one-loop conformal. Using (\ref{conf}), (\ref{conf4})
and (\ref{wzwrhat}), we find that the Einstein equations are satisfied 
for the affine-Sugawara constructions with 
$\del\Phi=0$ and central charge
\bs\be
c_G = \dim(g) - \frac{\a'}{2} H^2 + {\cal O}(\a'^2)
\ee\be
\qquad = \dim(g)(1-\frac{Q_g}{2k}) + {\cal O}(k^{-2})
\ee
\label{auxn10}
\es
which is in agreement with the exact 
form of the affine-Sugawara central charge
$c_g$ in (\ref{affb}).

\subsection{Strategy}

It is unlikely that the semiclassical form of the Einstein-Maxwell
formulation \cite{gme} of the VME can be identified directly with the
conventional Einstein equations (\ref{conf}) of the sigma model on a group
manifold. One reason is that the semiclassical inverse metrics
of the Einstein-Maxwell formulation
\be G^{ij}=\frac{1}{2k} e_a{}^i P^{ab} e_b{}^j + \ok \ee
are singular because $P$ is a projector (see
eqs (\ref{iii}) and (\ref{expan1})). On the other hand, we know
for example that the 
coset constructions are contained in both the
VME and the canonical stress tensors of the sigma model,
so the possibility remains open that
the Einstein-Maxwell form of the VME is in some sense dual to the Einstein
equations of the sigma model (see also the related remarks in
Section~7).

Our strategy here is a straightforward generalization of the VME
to the sigma model, following the relation of the general 
affine-Virasoro construction to the WZW model. In the operator algebra
of the general sigma model, we use the technique of Banks et
al. \cite{sig12} to study the general Virasoro construction
\bs\be
T = -\frac{L_{ij}}{\a '} \dif x^i \dif x^j 
+ {\cal O}(\a '^0)
 = -\frac{L^{ab}}{\a '} \Pi_a \Pi_b 
+ {\cal O}(\a '^0)
\label{genvira}
\ee\be
\bar{T} = -\frac{\bar{L}_{ij}}{\a '} \dbar x^i \dbar x^j 
 + {\cal O}(\a '^0)
  = -\frac{\bar{L}^{ab}}{\a '} \bar{\Pi}_a \bar{\Pi}_b 
+ {\cal O}(\a '^0)
\label{genvirb}
\ee\be
\dbar T = \dif \bar{T} = 0
\label{genvirc}
\ee\be
<T(z)T(w)> = 
\frac{c/2}{(z-w)^4} + \frac{2<T(w)>}{(z-w)^2} 
 + \frac{<\dif T(w)>}{(z-w)} + {\rm reg.}
\label{genvird}
\ee\be
<\bar{T}(\bar{z})\bar{T}(\bar{w})> = 
\frac{\bar{c}/2}{(\bar{z}-\bar{w})^4} + 
\frac{2<\bar{T}(\bar{w})>}{(\bar{z}-\bar{w})^2} 
 + \frac{<\dbar\bar{T}(\bar{w})>}{(\bar{z}-\bar{w})} + {\rm reg.}
\label{genvire}
\ee \label{genvir} 
\es
where the dilatonic contribution is included at 
${\cal O}(\a '^0)$
 and $L$ is a symmetric
second-rank spacetime tensor field (the inverse inertia tensor)
to be determined.

According to eq. (\ref{jjj19}), the stress tensors in (\ref{genvir})
reduce, at leading order in $\a'$, to the 
conventional stress tensors $T_G$, $\bar{T}_G$ of the general sigma
model when the inverse inertia tensors are taken as
\bs\be
L=\bar{L}=L_G
\ee\be
L_G^{ab}=\frac{G^{ab}}{2} + {\cal O}(\a')
\ee\es
where $G^{ab}$ is the inverse tangent space metric of the
sigma model action (\ref{sigmaact}). In this sense, the conventional
stress tensors of the sigma model generalize the affine-Sugawara 
stress tensors (\ref{affsug})
of the WZW model (see eq. (\ref{affa})),
and the general sigma model stress tensors (\ref{genvira}), (\ref{genvirb})
generalize the stress tensors (\ref{avc2})
 of the general affine-Virasoro construction.

\section{Classical preview of the construction}

The classical limit of the general construction (\ref{genvira}),
(\ref{genvirb})
can be studied with the classical equations of motion of the general
sigma model, which can be written in two equivalent forms
\be
\dbar \Pi_a + \bar{\Pi}_b \Pi_c \ho^{+bc}{}_a = 
\dif \bar{\Pi}_a + \Pi_b \bar{\Pi}_c \ho^{-bc}{}_a =0
\label{eqmot}
\ee
where $\Pi,\bar{\Pi}$ are defined in (\ref{defpi})  and $\ho^{\pm}$
are the generalized connections (\ref{tordefb}) with torsion.

One then finds that 
the classical stress tensors are holomorphic and antiholomorphic
respectively
\bs\be
T=-\frac{L^{ab}}{\a'} \Pi_a \Pi_b , \qquad \bar{T}=-\frac{\bar{L}^{ab}}{\a'}
\bar{\Pi}_a \bar{\Pi}_b 
\ee\be
\dbar T = \dif \bar{T} = 0
\ee
\label{tcla}
\es
iff the inverse inertia tensors are covariantly constant
\be
\delhp_i L^{ab} = \delhm_i \bar{L}^{ab} = 0
\label{lconst}
\ee
where $\delh^{\pm}$ are the generalized covariant derivatives (\ref{tordefa})
with torsion. Further discussion of these covariant-constancy conditions
is found in Sections~5.2 and especially~5.5, which places the relations
in the context of the theory of $G$-structures.

To study the classical Virasoro conditions, we introduce Poisson brackets
in Minkowski space. The Minkowski space action of the sigma model
\bs\be
S=\frac{1}{8\pi \a'} \int dt d\s (G_{ij}+B_{ij}) \dif_+ x^i \dif_- x^j
\label{auxn16}
\ee\be
\dif_{\pm}=\dif_t \pm \dif_{\s} 
\ee
\label{sigmaact2}
\es
is obtained from the Euclidean action (\ref{sigmaact}) by $z=e^{\t+i\s}$
and $t=i\t$. Then general Poisson brackets may be
computed from the fundamental brackets
\bs\be
[x^i(\s),p_j(\s')] = i \delta^i_j \delta(\s-\s')
\ee\be
p_i=\frac{1}{4\pi\a'}( G_{ij} \dif_t x^j - B_{ij} \dif_{\s} x^j)
\ee \label{fundalg} \es
where $p_i,i=1 \ldots \dim(M)$ are the canonical momenta.

We consider first the sigma-model currents
\bs\bea
J_{a}^\pm  & = &  G_{ab} e_{i}{}^b \dif_{\pm} x^i \\{}
&  = & e_a{}^i (4\pi\a' p_i + (B_{ij} \pm G_{ij} ) \dif_{\s} x^j )
\eea\es
which are the Minkowski space analogues of the Euclidean $\Pi,\bar{\Pi}$ in
(\ref{defpi}). After some algebra, we find that these currents satisfy
the general current algebra
\bs\be
[J^+_a(\s),J^+_b(\s')]=8\pi i \a' G_{ab} \dif_{\s} \delta(\s-\s') 
 + 4 \pi i \a' \delta(\s-\s') 
[  J^-_c \hat{\omega}^{+c}{}_{ab} -J^+_c \hat{\tau}^{+c}{}_{ab} ] 
\ee\be
[J^-_a(\s),J^-_b(\s')]=-8\pi i \a' G_{ab} \dif_{\s} \delta(\s-\s') 
 + 4 \pi i \a' \delta(\s-\s') [J^+_c \hat{\omega}^{-c}{}_{ab} 
 - J^-_c \hat{\tau}^{-c}{}_{ab} ] 
\ee
\vskip -.15in 
\be
[J^+_a(\s),J^-_b(\s')]=
  4 \pi i \a' \delta(\s-\s') [  \hat{\omega}^{+}_{ba}{}^c J^+_c
 -   \hat{\omega}^{-}_{ab}{}^{c} J^-_c ] 
\ee \label{alg} \es
where $\ho^{\pm}$ are the generalized connections (\ref{tordef}) with
torsion and the quantities
\be
(\hat{\tau}^{\pm})_{cab} \equiv \omega_{cab} + \o_{abc} + \o_{bca} 
\pm \frac{1}{2} H_{cab} 
\ee
are totally antisymmetric tensors on $M$. We have checked 
that the algebra (\ref{alg}) satisfies all Jacobi identities.

We have been unable to find the general current algebra (\ref{alg})
in the
literature of the sigma model, although Faddeev and Takhtajan \cite{sig14}
have given an example of the algebra in
the special case of the principal
chiral model, where
\be
G_{ab} = k \eta_{ab}, \qquad \o_{ab}{}^c = -\frac{1}{2\sqrt{\a'}}
f_{ab}{}^c, \qquad H_{ab}{}^c = 0.
\ee
For the special case of the WZW model, the algebra 
(\ref{alg}) is closely related
to the bracket form of affine $g \times g$: The currents $J^+$ already
satisfy the algebra of affine $g$ because $\hat{\o}^+=0$ in
this case, but the currents $J^-$ need a correction factor to complete
the algebra (see eq (\ref{auxn1b})). 
This subject is further discussed in Appendix
A, which also constructs a related set of non-local chiral currents in
the general sigma model.

We are now prepared to require that the chiral stress tensors
\be
T_{++}= \frac{1}{8\pi \a'} L^{ab} J^+_a J^+_b, \qquad
T_{--} = \frac{1}{8 \pi \a'} \bar{L}^{ab} J^-_a J^-_b
\ee
satisfy the commuting Virasoro algebras
\bs\be
[T_{\pm\pm}(\s),T_{\pm\pm}(\s')] = \pm i [T_{\pm\pm}(\s)
+ T_{\pm\pm}(\s')] \dif_{\s} \delta(\s-\s') 
\ee\be
[T_{++}(\s),T_{--}(\s')]=0.
\ee \label{viralg} \es
Using both the general current algebra (\ref{alg}) and the covariant constancy
(\ref{lconst}) of the inverse inertia tensors, we find 
after some algebra
that (\ref{viralg})
is satisfied iff
\be
L^{ab}= 2 L^{ac} G_{cd} L^{db} , \qquad \bar{L}^{ab} = 2 \bar{L}^{ac} G_{cd}
\bar{L}^{db}.
\label{lconst2}
\ee
These equations are the analogues on general manifolds of the high-level
or classical limit (\ref{expan2}) of the VME on group manifolds.

As discussed below, 
the classical relations
(\ref{lconst}) and (\ref{lconst2}) 
of this section will receive 
quantum corrections, e.g.
\bs\be
\delhp_i L^{ab} = {\cal O}(\a') 
\label{classa}
\ee\be
L^{ab} = 2 L^{ac} G_{cd} L^{db} + {\cal O}(\a')
\label{classb}
\ee 
\label{class} \es
although the generalized VME in (\ref{classb}) remains algebraic in $L$.

\section{Summary of the unified system}

We summarize here the results of our one-loop computation in the
general sigma model (\ref{sigmaact}), postponing details of the computation
until Section~6.

Including the one-loop dilatonic and counterterm contributions,
the holomorphic and antiholomorphic stress tensors
are 
\bs\be
T = - L^{ab} (\frac{\Pi_a \Pi_b}{\a '} + \frac{1}{2} \Pi_c \Pi_d 
H_{ae}{}^c H_b{}^{ed} ) + \dif (\Pi_a \Phi^a) + {\cal O}(\a')
\label{countera}
\ee\be
\bar{T} = - \bar{L}^{ab} (\frac{\bar{\Pi}_a \bar{\Pi}_b}{\a '} 
+ \frac{1}{2} \bar{\Pi}_c \bar{\Pi}_d 
H_{ae}{}^c H_b{}^{ed} ) + \dbar (\bar{\Pi}_a \bar{\Phi}^a) + {\cal O}(\a')
\label{counterb}
\ee
\vskip -.15in
\be
a,b = 1, \ldots, \dim(M)
\label{counterd} \ee
\label{counter}
\es
where $L^{ab}=L^{ba}$, $\bar{L}^{ab}=\bar{L}^{ba}$ are the inverse inertia
tensors and $\Pi_a$, $\bar{\Pi}_b$ are defined in (\ref{defpi}).
The second terms in $T$ and $\bar{T}$ are finite one-loop counterterms
which characterize our renormalization scheme. 
The quantities $\Phi^a$ and $\bar{\Phi}^a$ in 
(\ref{countera}), (\ref{counterb})
are called the dilaton vectors, and we will see below that the dilaton
vectors include the conventional dilaton as a special case.

For the holomorphic stress tensor $T$, we find the unified Einstein-Virasoro
master equation
\bs\be
L^{cd} \hat{R}^+_{acdb} + \delhp_a \Phi_b = {\cal O}(\a')
\label{eva}
\ee\be
\Phi_a = 2 L_a{}^b \Phi_b + \oa
\label{evn}
\ee\be
\delhp_i L^{ab} = {\cal O}(\a') 
\label{evb}
\ee\be
\eqalign{ L^{ab}  = & 2 L^{ac} G_{cd} L^{db} \cr
 & - \a' (L^{cd} L^{ef} H_{ce}{}^a H_{df}{}^b + 
 L^{cd} H_{ce}{}^f H_{df}^{(a} L^{b)e} ) \cr
& - \a' (L^{c(a} G^{b)d} \del_{[c} \Phi_{d]} ) + {\cal O}(\a'^2) \cr}
\label{evc}
\ee\be
c = 2 G_{ab} L^{ab} + 6 \a' (2 \Phi_a \Phi^a - \del_a \Phi^a)
 + {\cal O}(\a'^2)
\label{evd}
\ee
\label{ev}
\es
where the first line of (\ref{evc}) is the classical master equation in 
(\ref{lconst2}). 

The Virasoro conditions for the antiholomorphic stress tensor are quite
similar,
\bs\be
\bar{L}^{cd} \hat{R}^-_{acdb} + \delhm_a \bar{\Phi}_b = {\cal O}(\a')
\label{evba}
\ee\be
\bar{\Phi}_a = 2 \bar{L}_a{}^b \bar{\Phi}_b + \oa
\label{evbn}
\ee\be
\delhm_i \bar{L}^{ab} = {\cal O}(\a') 
\label{evbb}
\ee\be
\eqalign{ \bar{L}^{ab}  = & 2 \bar{L}^{ac} G_{cd} \bar{L}^{db} \cr
 & - \a' (\bar{L}^{cd} \bar{L}^{ef} H_{ce}{}^a H_{df}{}^b + 
 \bar{L}^{cd} H_{ce}{}^f H_{df}^{(a} \bar{L}^{b)e} ) \cr
& - \a' (\bar{L}^{c(a} G^{b)d} 
\del_{[c} \bar{\Phi}_{d]} ) + {\cal O}(\a'^2) \cr}
\label{evbc}
\ee\be
\bar{c} = 2 G_{ab} \bar{L}^{ab} + 6 \a' (2 \bar{\Phi}_a \bar{\Phi}^a 
- \del_a \bar{\Phi}^a)
 + {\cal O}(\a'^2)
\label{evbd}
\ee
\label{evball}
\es
and in fact follow from the unified system (\ref{ev}) for $T$, using the
symmetry $\Pi \leftrightarrow \bar{\Pi}$, $ H \leftrightarrow - H $ of
the sigma model action (\ref{sigmaact}).

In what follows, we refer to (\ref{eva}) and (\ref{evba}) as the 
generalized Einstein equations of the sigma model, and eqs.
(\ref{evn}) and (\ref{evbn}) are called the eigenvalue relations
of the dilaton vectors.
Equations (\ref{evc}), (\ref{evbc}) 
are called the generalized Virasoro master equations
(VMEs) of the sigma model. In
Section~5.7, we show that the central charges (\ref{evd}) and (\ref{evbd})
are consistent by Bianchi identities with the rest of the unified system.
The ${\cal O}(\a')$ corrections to the covariant-constancy
conditions (\ref{evb}) and (\ref{evbb}) can be computed in principle
from the solutions of the generalized VMEs.  

\vs .4cm
\noindent \underline{Some simple observations}
\vs .3cm 

\noindent
1. Algebraic form of the generalized VMEs.
In parallel
with the VME, the generalized VMEs (\ref{evc}) and (\ref{evbc}) are algebraic 
equations for $L$ and $\bar{L}$. This follows because any derivative of 
$L$ or $\bar{L}$ can be removed (see also Section~6) by using the
covariant-constancy conditions (\ref{evb}) and (\ref{evbb}). 


\vs .3cm

\noindent
2. Correspondence with the VME. The non-dilatonic terms of the
generalized
VMEs (\ref{evc}) and (\ref{evbc}) have exactly the form of the 
unimproved VME
(\ref{VME1}), after the covariant substitution
\be
f_{ab}{}^c \rightarrow \sqrt{\a'} H_{ab}{}^c
\label{auxn11}
\ee
for the general sigma model. This correspondence is the inverse of the
WZW datum in (\ref{wzwdata}),
\be H_{ab}{}^c = \frac{1}{\sqrt{\a'}} f_{ab}{}^c
\ee
which means that, for 
the special case of 
WZW, the non-dilatonic terms of the
generalized VMEs will reduce correctly to those of the 
unimproved VME. We return
to complete the WZW reduction in Section~5.2.

\vs .3cm

\noindent
3. Dilaton solution for the dilaton vector. According to the classical
limit (\ref{lconst2})
of the generalized VMEs, one solution of the eigenvalue relations
(\ref{evn}) and (\ref{evbn}) for the dilaton vectors is
\bs\be
 \Phi^a(\Phi)  \equiv 2 L^{ab} \del_b \Phi 
\label{eig1}
\ee\be
\bar{\Phi}^a(\Phi)  \equiv 2 \bar{L}^{ab} \del_b \Phi .
\label{eig2}
\ee \label{eig}
\es
In what follows, this solution is called the dilaton solution, and we
shall see in the following section that the scalar field $\Phi$ is in
fact the conventional dilaton of the sigma model.

\section{Properties of the unified system}

\subsection{The conventional stress tensors of the sigma model}

In this section, we check that the conventional stress tensors of the
sigma model are correctly included in the unified system.

In the full system, the conventional stress tensors 
$T_G$, $\bar{T}_G$ of the sigma
model correspond to the particular solution of the generalized VMEs
whose classical limit is 
\be
L^{ab} = \bar{L}^{ab} = L_G^{ab} = \frac{G^{ab}}{2} + \oa
\label{norms}
\ee
where $G^{ab}$ is the inverse of the metric in the sigma model action. 
The covariant-constancy conditions (\ref{evb}) and (\ref{evbb}) are 
trivially solved to this order because $\delh^{\pm}_i G_{ab}=0$.

To obtain the form of $T_G$ and $\bar{T}_G$ through one loop, 
we must also take
the dilaton solution (\ref{eig}) for the dilaton vectors, so that the
dilaton contributes to the system as
\bs\be
\Phi_a=\bar{\Phi}_a=\Phi_a^G = \del_a \Phi + \oa
\label{dnorma}
\ee\be
 \del_{[a} \Phi_{b]}^G = \oa .
\label{dnormb}
\ee\es
The relations (\ref{norms}) and (\ref{dnorma}) then tell us that the 
generalized Einstein equations (\ref{eva}) and (\ref{evba}) 
simplify to the conventional Einstein equations
\be
\hat{R}^{\pm}_{ab} - 2 \delh^{\pm}_a \delh^{\pm}_b \Phi = 
{\cal O}(\a ')
\ee
in agreement with equations (\ref{conf1}), (\ref{conf2}) and
(\ref{conf4}). Moreover, (\ref{dnormb}) tells us that the dilaton terms
do not contribute to the generalized VMEs in this case, and we may
easily obtain
\bs\be
L^{ab}=\bar{L}^{ab}=L^{ab}_G = \frac{G^{ab}}{2} -
\frac{\alpha'}{4} (H^2)^{ab} + \oaa 
\label{deflg}
\ee\be
\delh^{\pm}_i L^{ab}_G = -\frac{\alpha'}{4} 
 \delh^+_i (H^2)^{ab}  + \oaa
\ee\be
T_G(\Phi) = -\frac{G^{ab}}{2\a'}\Pi_a \Pi_b  + \dif^2 \Phi + \oa
\label{zz1}
\ee \be
\bar{T}_G(\Phi)  = -\frac{G^{ab}}{2\a'} \bar{\Pi}_a \bar{\Pi}_b +
\dbar^2 \Phi + \oa
\label{zzz1}
\ee\es
by solving the generalized VMEs through the indicated order. In this
case, the stress tensor counterterms in (\ref{countera}), (\ref{counterb})
cancel against the $\oa$ correction to $L_G$,
and (\ref{zz1}), (\ref{zzz1}) are consistent with (\ref{jjj19}).
In what follows, the stress
tensors $T_G(\Phi)$ and $\bar{T}_G(\Phi)$
 are called the conventional stress
tensors of the sigma model.

To complete the check, we evaluate the central charges
$c=\bar{c}=c_G(\Phi)$ in this case,
\bs\be
c_G(\Phi)=2 G_{ab} (\frac{G^{ab}}{2} -\frac{\a'}{4}  (H^2)^{ab} ) 
+ 6 \a' (2 |\del \Phi|^2 - \del^2 \Phi) + \oaa 
\label{ccca}
\ee\be
\qquad = \dim(M)  + 3 \a' (4 |\del \Phi|^2 - 2 \del^2 \Phi -
\frac{1}{6} H^2 ) + \oaa
\label{cccb}
\ee\be
\qquad = \dim(M)  + 3 \a' (4 |\del \Phi|^2 - 4 \del^2 \Phi +R +
\frac{1}{12} H^2 ) + \oaa
\label{cccc}
\ee \label{ccc} \es
which agrees with the conventional central charge in (\ref{conf3}). To
obtain the usual form in (\ref{cccc}), we used the conventional 
Einstein equations (\ref{conf1}) in the form $R=2\del^2 \Phi -\frac{1}{4}
H^2$. 

\vs .4cm
\noindent \underline{Dilaton-vector Einstein equations}
\vs .3cm 

For completeness, we also note the form of the system for
$L=\bar{L}=L_G$ with general dilaton vectors $\Phi^G_a$, $\bar{\Phi}^G_a$.
The dilaton vector terms still do not contribute to the generalized 
VMEs, whose solution is still (\ref{deflg}).
For the holomorphic system, one obtains
\bs\be
T_G(\Phi_a) = -\frac{G^{ab}}{2\a'} \Pi_a \Pi_b + \dif(\Pi^a \Phi_a^G) +\oa
\label{qqa}
\ee\be
c_G(\Phi_a) = \dim(M) + 3\a' (4 \Phi^G_a \Phi_G^a - 4 \del_a \Phi^a_G + R
 + \frac{1}{12} H^2 ) + \oaa
\label{qqb}
\ee\be
\hat{R}^+_{ab} - 2 \delhp_a \Phi^G_b = \oa
\label{qqc}
\ee \label{qq} \es
where $\Phi^a_G$ is unrestricted because its eigenvalue equation is trivial.
The antiholomorphic system in this case is similarly obtained with 
$\Pi_a \rightarrow \bar{\Pi}_a$, $+ \rightarrow -$ and $\Phi^G_a 
\rightarrow \bar{\Phi}^G_a$. In what follows, the stress tensor 
$T_G(\Phi_a)$ in (\ref{qqa})
will be called the the conventional stress tensor with general dilaton
vector, and the relations in (\ref{qqc}) will be
called the dilaton-vector Einstein equations of the sigma model.

\subsection{WZW and the improved VME}

In this section we check that, for the special case of WZW, the unified
system reduces to a holomorphic and an antiholomorphic copy of the improved
VME (\ref{VME1}), where the improvement vectors $D,\bar{D}$ are constructed
from the general dilaton vectors.

We recall first from Section~4 that the non-dilatonic terms 
of the generalized VMEs 
reduce correctly to those of the VME,
\bs\bea
L^{ab} &  = &  (\mbox{{\rm usual $L^2$ and $L^2f^2$ terms}}) + 
\mbox{{\rm dilaton vector term}} + \oaa 
\\{}
\bar{L}^{ab} & =  & (\mbox{{\rm usual $\bar{L}^2$ and $\bar{L}^2f^2$ terms}}) + 
\mbox{{\rm dilaton vector term}} + \oaa 
\eea\es
because $H=\frac{1}{\sqrt{\a'}} f$ in this case.

To study the dilaton vector terms, we recall first that the generalized Riemann
tensors $\hat{R}^{\pm}_{abcd}$ vanish for WZW, so the generalized Einstein
equations (\ref{eva}) and (\ref{evba}) simplify to
\be \delhp_a \Phi_b = \oa, \qquad \delhm_a \bar{\Phi}_b = \oa .
\label{auxn7}
\ee
These relations and the WZW datum (\ref{wzwdata}) for $H$ can be used
to evaluate the (ordinary) covariant derivatives of the dilaton vectors
\be
\del_a \Phi_b = \frac{1}{2\sqrt{\a'}} f_{ab}{}^c \Phi_c + \oa,
\qquad
\del_a \bar{\Phi}_b =- \frac{1}{2\sqrt{\a'}} f_{ab}{}^c \bar{\Phi}_c + \oa,
\label{aux6}
\ee
so that the generalized VMEs (\ref{evc}) and (\ref{evbc}) have the
form
\bs\bea
L^{ab} &  = &  (\mbox{{\rm usual $L^2$ and $L^2f^2$ terms}}) + 
\sqrt{\a'} f_{cd}{}^{(a} L^{b)c} \Phi^d + \oaa
\\{}
\bar{L}^{ab} &  = &  (\mbox{{\rm usual $\bar{L}^2$ and $\bar{L}^2f^2$ terms}}) -
\sqrt{\a'} f_{cd}{}^{(a} \bar{L}^{b)c} \bar{\Phi}^d + \oaa
\eea\es
when the sigma model is taken as WZW. Comparing with the improved VME
(\ref{VMEall}), one is tempted to identify the constant improvement vectors
$D^a$, $\bar{D}^a$ as proportional by constants to the dilaton 
vectors $\Phi^a$, $\bar{\Phi}^a$. The full story is slightly more
involved however, because we have not yet studied the spacetime
dependence of the dilaton vectors.

\vs .4cm
\noindent \underline{Constant quantities on the group manifold}
\vs .3cm 

To study the coordinate dependence of $\Phi^a$ and $\bar{\Phi}^a$
for WZW, we return to solve
the generalized Einstein equations 
(\ref{auxn7}) using the WZW data
\be \hat{\omega}^+{}_{ab}{}^c  =0,\qquad 
\hat{\omega}^-{}_{ab}{}^c  = -\frac{1}{\sqrt{\a'}} f_{ab}{}^c.
\label{auxn8}
\ee
At leading order, the general solution of these equations can be 
written as conservation laws
\bs\be
\dif_i D^a = \dif_i \bar{D}^a = 0 
\label{consa}
\ee\be
D^a \equiv - \sqrt{\a'} \Phi^a = {\rm constant}
\label{consb}
\ee\be
\bar{D}^a \equiv  \sqrt{\a'} \bar{\Phi}^b \Omega_{b}{}^a
 = {\rm constant}
\label{consc}
\ee\label{cons} \es
Here $\Omega_a{}^b$ is the adjoint action of $g$, whose differential
equation (\ref{aux4}) was used to obtain the solution
(\ref{consc}), and we
have chosen the multiplicative constants in (\ref{consb}), (\ref{consc})
so that the constant quantities 
$D,\bar{D}$ will turn out to be the improvement vectors.

One way to understand why $\bar{\Phi}^a$ is not a constant is that 
$\bar{L}^{ab}$ is not a constant either. To see this, we similarly solve
the covariant-constancy conditions (\ref{evb}), (\ref{evbb}) for
the inverse inertia tensors, using the WZW data (\ref{auxn8}) again.
At leading order the solution is
\bs\be
\dif_i L^{ab} = \dif_i \bar{L}'^{ab} = 0 
\ee\be
L^{ab} = {\rm constant}
\ee\be
\bar{L}'^{ab} = \bar{L}^{cd} \Omega_c{}^a \Omega_d{}^b = {\rm constant}
\ee\es
where $\Omega_a{}^b$ is again the adjoint action of $g$.

This completes the picture for the generalized VMEs, and we see that
both can be written in terms of the constant 
quantities\footnote{In the high-level expansion \cite{hl,rev} of the VME,
it is known that $L^{ab}={\cal O}(k^{-1})$ and the $L^2 f^2$ terms of the
VME are ${\cal O}(k^{-2})$, which corresponds to one-loop at the sigma model
level. For correspondence with the loop expansion,
the improvement 
term $fLD$ must then also be ${\cal O}(k^{-2})$, which says that 
$D^a={\cal O}(k^{-1})$. To obtain improvement vectors $D^a={\cal O}(k^0)$,
one must allow dilaton-vector contributions at the tree level (see also
the remarks in Section~7).},
\bs\be
L^{ab} = (\mbox{{\rm usual $L^2$ and $L^2f^2$ terms}}) -
  f_{ce}{}^{(a} L^{b)c} D^e + {\cal O}(k^{-3})
\label{chka}
\ee\be
\bar{L}'^{ab} = (\mbox{{\rm usual $\bar{L}'^2$ and $\bar{L}'^2f^2$ terms}}) -
 f_{ce}{}^{(a} \bar{L}'^{b)c} \bar{D}^e + {\cal O}(k^{-3})
\label{chkb}
\ee 
\label{chk}
\es
because $\Omega$ is an inner automorphism of $g$ and the VME is covariant 
under transformations in $\rm{Aut}(g)$ (see Section~2.1). The result
(\ref{chk}) is exactly two copies of the improved VME (\ref{VME1}), with
the improvement vectors $D,\bar{D}$ constructed from the dilaton vector
in (\ref{consb}), (\ref{consc}).

The reason for the extra automorphism in the antiholomorphic system
is in fact clear
from our starting point. Both $\Pi$ and $\bar{\Pi}$ are defined with
the vielbein $e$ on $M$, which we have identified with the left-invariant
vielbein on $G$, but the antiholomorphic WZW current $\bar{J}$
involves instead the right-invariant vielbein $\bar{e}$.
This difference is responsible for the extra factor of the adjoint
action $\Omega$ in the antiholomorphic part of the WZW data (\ref{auxn1}),
\be J_a=\frac{i}{\sqrt{\a'}} \Pi_a, \qquad
\bar{J}_a = -\frac{i}{\sqrt{\a'}} (\Omega^{-1})_a{}^b \bar{\Pi}_b.
\label{jjbar} \ee
Indeed, using this data and the identifications (\ref{consb}), (\ref{consc})
of $D$ and $\bar{D}$, we find that the sigma model stress tensors 
(\ref{countera}), (\ref{counterb}) reduce to the form
\bs\be
T=L^{ab} J_a J_b + iD^a \dif J_a + (\mbox{{\rm $L$-counterterm}}) + \oa
\ee\be
\bar{T}=
\bar{L}'^{ab} \bar{J}_a 
\bar{J}_b + i\bar{D}^a 
\dif \bar{J}_a + (\mbox{{\rm $\bar{L}'$-counterterm}}) + \oa
\ee\es
which is nothing but two copies of the general affine-Virasoro construction.

Next, we consider the central charges (\ref{evd}) and (\ref{evbd}), using
the relations
\be
\del_a \Phi^a = \del_a \bar{\Phi}^a = \oa
\ee
which follow for WZW from eq. (\ref{aux6}). Then the central charges reduce
to
\bs\be
c=2 G_{ab} (L^{ab} + 6 D^a D^b) + {\cal O}(k^{-2})
\label{redca}
\ee\be
\bar{c}=2 G_{ab} (\bar{L}'^{ab} + 6 \bar{D}^a \bar{D}^b)+ {\cal O}(k^{-2})
\label{redcb}
\ee\es
in agreement with the central charge (\ref{VME3}) of the improved VME. The
central charges can be taken equal $\bar{c}=c$ by choosing $\bar{L}'=L$ and
$\bar{D}=D$.

We finally note that the eigenvalue relations (\ref{evn}), (\ref{evbn})
 of the dilaton
vectors can be rewritten with (\ref{cons}) as
\bs\be
2 L^{ab} G_{bc} D^c = D^a + {\cal O}(k^{-2})
\label{eeig}
\ee\be
2 (\bar{L}')^{ab} G_{bc} \bar{D}^c = \bar{D}^a +{\cal O}(k^{-2})
\ee \label{z4} \es
which are recognized as 
the leading terms of two copies of the exact eigenvalue
relation (\ref{VME2}) of the improved VME.
This completes the one-loop check of the unified system against the 
improved VME. 

\vs .4cm
\noindent \underline{Dilaton deformations in WZW theory}
\vs .3cm 

Except for their eigenvalue relations, the improvement vectors $D^a$, $\bar{D}^a$
of the VME are quite general, and are in correspondence with the general
dilaton vectors $\Phi^a$, $\bar{\Phi}^a$ of the unified system with WZW
background. In particular, it is well known \cite{impro}
 that the improvement vector
$D_g^a$ for the affine-Sugawara construction $L_g$ is completely general,
\be
D_g^a = \mbox{\rm arbitrary constants}
\label{zzzz}
\ee
because the eigenvalue relation (\ref{VME2}) is an identity in this case.

On the other hand, the dilaton solution 
$\Phi_a(\Phi)$ in (\ref{eig}) for the dilaton vectors 
provides only a restricted subset of these deformations, which in fact may
be more restricted than it appears at first sight.
As an
example, consider the possible dilaton deformations
\be D^a_g(\Phi)  = -\sqrt{\a '} \Phi^a_g = -\sqrt{\a '} \del_a \Phi
\ee 
of the affine-Sugawara construction $L_g$. In this case we have from 
(\ref{aux6}) that 
\bs\be
\oa = [\del_a,\del_b] \Phi = \frac{1}{\sqrt{\a'}} f_{ab}{}^c 
\Phi^g_c = \frac{1}{\sqrt{\a'}} f_{ab}{}^c \del_c \Phi
\ee\be
\rightarrow \Phi={\rm constant}, \qquad D^a_g(\Phi)=0
\ee\es
so that, in contrast to (\ref{zzzz}), no dilaton
deformations are allowed for the affine-Sugawara constructions.

\vs .4cm
\noindent \underline{Exact relations in WZW theory?}
\vs .3cm 

Beyond one loop, it should be emphasized that
the improved VME (\ref{VME1}), (\ref{VME2})
 and central charge (\ref{VME3}) are exact
to all orders in this case
and the quantities $L^{ab}$, $L'^{ab}$, $D^a$ and $\bar{D}^a$
are constant to all orders. It follows that the covariant-constancy
conditions
\bs\be
\delhp_i L^{ab} = \delhm_i \bar{L}^{ab} = 0
\label{cocoa} \ee\be
\delhp_i D^a = \delhm_i (\bar{D}^b (\Omega^{-1})_b{}^a ) =0
\label{cocob} \ee \label{coco} \es
are exact to all orders in the general affine-Virasoro construction.
The one-loop form of (\ref{cocoa}), e.g.
$\delhp_i L^{ab}=\oaa$, also follows from the one-loop master equations
(\ref{chk}). This tells us that it is reasonable 
to assume that a higher-loop
renormalization scheme can be found 
which preserves the VME and the relations
(\ref{coco}) 
at arbitrary loop-order when the sigma model is taken as WZW.

If we further assume that our correspondence (\ref{consb}), (\ref{consc}),
\be
D^a = -  \sqrt{\a'} \Phi^a, \qquad \bar{D}^a = \sqrt{\a'} 
\bar{\Phi}^b  \Omega_b{}^a \label{dexa}
\ee
is also exact, we are led to write down the relations
\bs\be
\delhp_i \Phi_a = \delhm_i \bar{\Phi}_a = 0 
\label{rela} \ee\be
\Phi^a( 2 G_{ab} L^{be} + f_{ab}{}^d L^{bc} f_{cd}{}^e) = \Phi^e 
\label{relb} \ee\be
\bar{\Phi}^a( 2 G_{ab} \bar{L}^{be} + f_{ab}{}^d \bar{L}^{bc}
 f_{cd}{}^e) = \bar{\Phi}^e 
\label{relc} \ee \label{rel} \es
as candidates for exact eigenvalue
relations in WZW theory. Here, (\ref{rela})
follows from (\ref{cocob}) and (\ref{dexa}), 
while (\ref{relb}), (\ref{relc}) are transcriptions of the exact
eigenvalue condition (\ref{VME2}) using (\ref{dexa}). The 
candidate relations
(\ref{rel}) take the simpler form
\bs\be
\delh^{\pm}_i \del_j \Phi=0  \label{fin1a}
\ee\be
\del^a \Phi (2G_{ab} L_g^{be} + f_{ab}{}^d L_g^{bc} f_{cd}{}^e)=
\del^e \Phi \label{fin1b}
\ee \label{fin1} \es
for the dilaton of the conventional stress tensors (\ref{zz1}),
where $L_g^{ab}$ is the inverse
inertia tensor 
(\ref{affa}) of the affine-Sugawara construction. 
In this case, the literature contains some supporting evidence:
The analysis of 
\cite{mukhi,sig13} shows that there is a renormalization scheme in which
(\ref{fin1a}) holds, and, if in such a scheme the VME is also exact,
then relation (\ref{fin1b}) is a consequence of the exact form of
$L_g$ in (\ref{afff}).

\subsection{Exact relations in the general sigma model?}

Although our unified system has been computed only through one loop,
we saw in Sections~5.1 and~5.2 that some of the components of the system
are both
\begin{itemize}
\item in one-loop agreement with the conventional stress tensors of the 
general sigma model.
\item exact for the general affine-Virasoro construction.
\end{itemize}
We collect those components here, since,
with an appropriate renormalization scheme, they are candidates for
exact relations in the general sigma model.

Among these candidates, we mention first the generalized VME and its
central charge
\bs\be
\eqalign{ L^{ab}  = & 2 L^{ac} G_{cd} L^{db} \cr
 & - \a' (L^{cd} L^{ef} H_{ce}{}^a H_{df}{}^b + 
 L^{cd} H_{ce}{}^f H_{df}^{(a} L^{b)e} ) \cr
& - \a' (L^{c(a} G^{b)d} \del_{[c} \Phi_{d]} )  \cr}
\label{exeva}
\ee\be
c = 2 G_{ab} L^{ab} + 6 \a' (2 \Phi_a \Phi^a - \del_a \Phi^a)
\label{exevb}
\ee \label{exev} \es
and the corresponding antiholomorphic copy in (\ref{evball}). As discussed
in the previous sections, these relations are in one-loop agreement with the
conventional stress tensors of the general model, and, under the 
identification (\ref{dexa}), they are exact to all orders in the
general affine-Virasoro construction. We remark in particular that the
(one-loop) generalized VMEs contain no terms
proportional to the generalized Riemann tensors $\hat{R}^{\pm}$.
Such terms would not affect the check against the general affine-Virasoro
construction, but would invalidate (\ref{exeva}), (\ref{exevb}) as
candidates for exact relations in the sigma model.

To this list, we should add the eigenvalue
relations for the dilaton vectors
\bs\be
\Phi^a (2 G_{ab} L^{be} + \a' H_{ab}{}^d L^{bc} H_{cd}{}^e) = 
\Phi^e \label{dileiga}
\ee\be
\bar{\Phi}^a (2 G_{ab} \bar{L}^{be} + \a' H_{ab}{}^d 
\bar{L}^{bc} H_{cd}{}^e) = 
\bar{\Phi}^e 
\label{dileigb} \ee \label{dileig} \es
which we have constructed from the exact eigenvalue relations (\ref{VME2})
using the correspondence (\ref{dexa}) and the covariant substitution
$f_{ab}{}^c \rightarrow \sqrt{\a'} H_{ab}{}^c$ in (\ref{auxn11}).
These relations generalize the one-loop eigenvalue relations 
(\ref{evn}), (\ref{evbn})
and are in all-order agreement with the general affine-Virasoro construction.

The eigenvalue relations (\ref{dileig}) take a simpler form for the
conventional stress tensors $T_G(\Phi)$
of the general sigma model,
\be \del^a \Phi(2 G_{ab} L_G^{be} + \a' 
H_{ab}{}^d L_G^{bc} H_{cd}{}^e) = \del^e \Phi
\label{dileig2}
\ee
where the inverse inertia tensor $L_G$ is given in
(\ref{deflg}). Using the explicit form of $L_G$, we can easily
check that 
\be
2G_{ab} L_G^{be} + \a' H_{ab}{}^d L_G^{bc} H_{cd}{}^e = \delta_a{}^e
+ \oaa
\ee
so the candidate relation (\ref{dileig2}) is verified at two loops in the
form $\del^e\Phi=\del^e \Phi + \oaa$.

Investigation of the relations (\ref{exev}) and (\ref{dileig}) at higher
order is an important open problem in ICFT since such relations would,
through the Bianchi identities, place strong constraints on the higher-order
form of the generalized and conventional Einstein equations of the
general sigma model.

\subsection{Semiclassical properties of the system}

The generalized VME (\ref{evc}) can be written in the form
\bs\be L_a{}^b = 2 L_a{}^c L_c{}^b +
\a' X(L)_a{}^b + \oaa 
\ee\be
\delhp_i L_a{}^b = \oa
\ee\es
where $L_a{}^b=L^b{}_a=G_{ac}L^{cb}$ and the matrix
$X_a{}^b$ is easily read from
(\ref{evc}). The solutions of this system may be studied as power series
in the string tension $\a'$
\bs\be
L_a{}^b = \frac{1}{2} P_a{}^b + \a' \Delta_a{}^b + \oaa 
\label{psera}
\ee\be
P_a{}^c P_c{}^b = P_a{}^b, \qquad \delhp_i P_a{}^b =0
\label{pserb} \ee\be
\Delta_a{}^b = P_a{}^c \Delta_c{}^b + \Delta_a{}^c P_c{}^b 
+ X(P/2)_a{}^b
\label{pserc} \ee\be
\delhp_i L_a{}^b = \a' \delhp_i \Delta_a{}^b + \oaa
\label{pserd} \ee\label{pser} \es
where $P$ is a covariantly constant projector and $\Delta$, which must
solve the linear equation (\ref{pserc}), is the $\oa$ correction to $L$.
The same forms are found for the antiholomorphic subsystem with
$L\rightarrow \bar{L}$ and $P\rightarrow \bar{P}$, where $\bar{P}$ is
a covariantly constant projector which satisfies $\bar{P}^2=\bar{P}$ and
$\delhm_i \bar{P}=0$.

The zeroth-order generalized Einstein and dilaton vector equations are then
\bs\be
P^{cd} \hat{R}^+_{acdb} + 2\delhp_a 
\Phi^{(0)}_b = \bar{P}^{cd} \hat{R}^-_{acdb} + 2 \delhm_a 
\bar{\Phi}^{(0)}_b=0
\label{zeva} 
\ee\be
\Phi_a^{(0)}=P_a{}^{b} \Phi_b^{(0)}, \qquad
\bar{\Phi}_a^{(0)}=\bar{P}_a{}^{b}  \bar{\Phi}_b^{(0)}, 
\label{zevb}
\ee\es
where 
$\Phi_a^{(0)}$ and
$\bar{\Phi}_a^{(0)}$ are the zeroth-order form of the dilaton vectors and
$P$, $\bar{P}$ are the covariantly constant projectors.

The classical projector $P$ and the linear equation (\ref{pserc}) are in
close correspondence with the high-level expansion \cite{hl,rev} of the
VME, so we know that (\ref{pserc}) will place further restrictions
on the allowed projectors and that a general solution of (\ref{pserc}) 
is unlikely. (See however the known solutions \cite{rev} of the VME
on group manifolds 
and the particular solution $L_G$ in (\ref{deflg}) on any manifold.)

On the other hand, we may easily determine $L_a{}^a=\tr(L)$ 
by studying $\tr(\Delta)$ and $\tr(P\Delta)$, using (\ref{pserc}).
This gives the expansions 
\bs\be
2L_a{}^a={\rm rank}(P) + \a' (P_a{}^b P_c{}^e
(P_d{}^f-\frac{3}{2} \delta_d^f))H^{acd} H_{bef} + \oaa
\label{ltracea}
\ee\be
{\rm rank}(P) = 4(L_a{}^b L_b{}^a - 2 \a' L_a{}^b
L_c{}^e (L_d{}^f - \delta_d^f) H^{acd} H_{bef}) + \oaa
\label{ltraceb}
\ee\label{ltrace} \es
and similarly for $L\rightarrow \bar{L}$, $P\rightarrow \bar{P}$. The
relation (\ref{ltracea}) generalizes the known \cite{hl} high-level expansion
of the central charge in the general affine-Virasoro construction, and,
using the generalized VME, the relation (\ref{ltraceb}) follows
from (\ref{ltracea}). Both relations will be useful below.

\subsection{Integrability conditions and $G$-structures}

The inverse inertia tensors $L^{ab}$ and $\bar{L}^{ab}$ are 
second-rank symmetric spacetime 
tensors, and we know that their associated projectors
$P$ and $\bar{P}$ are covariantly constant
\be
\delhp_i P_a{}^b = \delhm_i \bar{P}_a{}^b = 0
\label{pcon}
\ee
Operating with a second covariant derivative, we find that the 
integrability conditions
\bs\be
\hat{R}^+_{cd}{}^{ae} P_e{}^b + \hat{R}^+_{cd}{}^{be} P_e{}^a = 0
\label{aux9a}
\ee\be
\hat{R}^-_{cd}{}^{ae} \bar{P}_e{}^b + \hat{R}^-_{cd}{}^{be} \bar{P}_e{}^a = 0
\label{aux9b}
\ee \label{aux9} \es
follow as necessary conditions for the existence of solutions to (\ref{pcon}).
In the following, we give two applications of the integrability conditions.

\vs .4cm
\noindent \underline{Ricci form of the generalized Einstein equations}
\vs .3cm 

Using (\ref{pcon}) and the integrability conditions (\ref{aux9}), the
zeroth-order generalized Einstein equations (\ref{zeva}) are
easily written in the Ricci form
\be
(\hat{R}^+_{ab} - 2 \delhp_a \Phi_b^{(0)} ) P^{bc} = 
(\hat{R}^-_{ab} - 2 \delhm_a \bar{\Phi}_b^{(0)} ) \bar{P}^{bc} =0
\label{auxn12}
\ee
where the left factors have the same form as the dilaton-vector
Einstein equations (\ref{qqc}).
Moreover, the steps
followed to obtain the Ricci form are reversible, so that the
Ricci form (\ref{auxn12}) and the generalized Einstein equations
(\ref{zeva}) are equivalent when $P$ and $\bar{P}$ are covariantly 
constant.

\vs .4cm
\noindent \underline{Covariant derivative of $\tr(L)$}
\vs .3cm 

As a second application of the integrability conditions, we will
verify the relations
\bs\be
\delhp_i L_a{}^a = -\frac{3}{2} \a' P_c{}^b H_{bed} \hat{R}_i^{+edc} +
\oaa 
\label{dtracea} 
\ee\be
\delhm_i \bar{L}_a{}^a = +\frac{3}{2} \a' \bar{P}_c{}^b H_{bed} 
\hat{R}_i^{-edc} + \oaa 
\label{dtraceb} 
\ee\label{dtrace} \es
giving details only for (\ref{dtracea}). These relations will be useful
below in checking that the central charges are covariantly constant.

For the proof we note first that, 
according to (\ref{ltracea}) and (\ref{pcon}), the relation (\ref{dtracea})
is equivalent to the relation
\be
\delhp_i (P_a{}^b P_c{}^e (P_d{}^f - \frac{3}{2} \delta_d{}^f)
H^{acd} H_{bef} ) = -3 P_c{}^b H_{bed} \hat{R}_i^{+edc}
\label{aux10}
\ee
whose left and right sides are respectively cubic and linear in the
projector $P$. To establish (\ref{aux10}), we compute directly using
$\delhp_i P=0$, $P^2=P$, the integrability conditions (\ref{aux9a}) and
the $(+)$ form of the Bianchi identities 
\be
\delh^{\pm}_i H_{abc} = \pm (\hat{R}^{\pm}_{iabc}+
\hat{R}^{\pm}_{icab}+ \hat{R}^{\pm}_{ibca})
\label{cyclic}
\ee
which follow from the explicit form of $\hat{R}^{\pm}_{abcd}$ in
(\ref{tordeff}).

As an example, we follow that term 
in  the differentiation (\ref{aux10}) which corresponds to the
first term in (\ref{cyclic}),
\bs\be
2P_a{}^b P_c{}^e (P_d{}^f - \frac{3}{2} \delta_d^f) H_{bef} 
\hat{R}_i^{+acd}
\label{aux11a} \ee\be
= -2P_a{}^b (P_d{}^f - \frac{3}{2} \delta_d^f) H_{bef} 
\hat{R}_i^{+adc} P_c{}^e
\label{aux11b} \ee\be
=2P_a{}^b P_c{}^d (P_d{}^f - \frac{3}{2} \delta_d^f) H_{bef} 
\hat{R}_i^{+aec}
\label{aux11c} \ee\be
=-P_a{}^b P_c{}^f H_{bef} 
\hat{R}_i^{+aec}
\label{aux11d} \ee \label{aux11}\es
where we used the integrability conditions (\ref{aux9a}) to obtain
(\ref{aux11c}) and $P^2=P$ to obtain (\ref{aux11d}).
Following similar steps for the other terms, we find
that the left side of (\ref{aux10}) 
is reduced to a quadratic form in $P$, and that the
quadratic terms in fact cancel, verifying the right side of (\ref{aux10})
and hence (\ref{dtracea}). The relation (\ref{dtraceb}) follows
similarly from the $(-)$ form of the Bianchi identities in 
(\ref{cyclic}).

\vs .4cm
\noindent \underline{Theory of $G$-structures}
\vs .3cm 

On any manifold, there is always at least one solution to the
covariant-constancy conditions (\ref{pcon}) and 
their integrability conditions (\ref{aux9}), namely
\bs\be
P^{ab}=\bar{P}^{ab}=G^{ab}
\ee\be
\hat{R}^{\pm ab}_{cd} +\hat{R}^{\pm ba}_{cd} = 0 
\ee\be
L^{ab}=\bar{L}^{ab}=L_G^{ab} = 
\frac{G^{ab}}{2} + \oa
\ee\es
where $G_{ab}$ is the metric of the sigma model action. This solution 
is the classical limit of the conventional sigma model stress tensor,
as discussed in Section~2.2. For WZW, the integrability conditions 
(\ref{aux9}) are also trivially satisfied (because 
$\hat{R}^{\pm}_{abcd}=0$) and the general solutions of the
covariant-constancy conditions were obtained for this case in
Section~5.2.

In general we are interested in the classification of manifolds with at
least one more solution $P^{ab}$, 
beyond $G^{ab}$. This is a problem in the theory of $G$-structures 
\cite{hol12}, which includes the study 
of all possible covariantly constant objects on 
arbitrary manifolds. Physical examples include the case of
Calabi-Yau manifolds \cite{gsw,hol2}, which
are characterized by the presence of a covariantly constant spinor,
and supersymmetric sigma models \cite{hol11} with additional 
conserved currents, which are characterized by the presence of
covariantly closed differential forms in target space. 
Here, we are looking for manifolds with one or more
 covariantly constant projectors
$P^{ab}$ beyond $G^{ab}$.

To get an idea of the properties of such manifolds,
we start by examining the covariant-constancy conditions
(\ref{pcon}). The equation $\delhp_i P_a{}^b=0$
is a first-order linear differential equation for $P_a{}^b$. This means
that $P_a{}^b$ is completely determined by its value at one point,
say $P_a{}^b(x_0)$. The value of $P_a{}^b$ at an arbitrary point $x$ can
then be obtained by integrating $\delhp_i P_a{}^b=0$ along a curve
$\gamma$ from $x_0$ to $x$, or, equivalently, by parallel transport of
$P_a{}^b$ along $\gamma$ with respect to the connection $\delhp$. For
consistency, the result should not depend on the choice of curve $\gamma$.
Two different curves $\gamma_1$ and $\gamma_2$ give the same result if
and only if we recover the original value $P_a{}^b(x_0)$ 
after parallel transport 
along the closed curve $\zeta$ which begins and ends at $x_0$, and
which is given by the union of $\gamma_1$ and $-\gamma_2$. 
The parallel transport of any object along a closed curve $\zeta$ can
by definition be obtained from the holonomy of $\zeta$, which is a matrix
$M(\zeta)\in O(d)$. In the case of $P_a{}^b(x_0)$, its parallel
transport along $\zeta$ is given in terms of $M$ by
\be P_a{}^b(x_0) \stackrel{\zeta}{\rightarrow} M(\zeta)_a{}^c
(M(\zeta)^t)_d{}^b P_c{}^d(x_0) \ee
where $M(\zeta)^t$ denotes the transpose of $M(\zeta)$.
Thus, the consistency requirement that (along each closed curve $\zeta$
starting and ending at $x_0$) $P_a{}^b(x_0)$ is parallel transported
back to itself now reads that for all $\zeta$
\be
P_a{}^b(x_0) M(\zeta)_b{}^e = M(\zeta)_a{}^c P_c{}^e(x_0).
\label{invar}
\ee
The set of all matrices $M(\zeta)$ from a group, the holonomy group 
[45--47,43,48]
of the target-space manifold $M$, and (\ref{invar}) tells us that $\ker(P)$
is an invariant subspace of the representation $M(\zeta)_a{}^c$ of the
holonomy group, which must therefore be reducible (except for the
trivial cases when $P^{ab}=0$ or $P^{ab}=G^{ab}$). 

In the case that $\delhp$ is torsion-free (i.e. $H=0$), it is known
from the de Rham global splitting theorem \cite{hol7}
that reducibility of the holonomy implies 
that the manifold must be a direct product of a certain number
of submanifolds with the direct
product metric, each of which has irreducible holonomy,
and all possible irreducible holonomies of each submanifold have been
completely classified [50--53].
As a result, the torsion-free case contains only 
trivial
solutions of the unified system where
the sigma model is a direct product of a conformal and a non-conformal
field theory (see also Section~5.8).
More interesting is the broader situation where the antisymmetric
tensor, while still closed,
is nonvanishing. In that case it is an unsolved problem which
manifolds have reducible holonomy, except when the holonomy is trivial.
In the latter case, the manifold must be a group manifold, and the
sigma model must be a WZW model. A derivation of this fact
is included in Appendix~A.

To summarize,
the classification of manifolds with reducible holonomy is centrally
relevant for the solutions of the unified system which are solutions
for generic values of $\a'$. For any such manifold the allowed 
projectors $P$ are in one-to-one correspondence with the projectors
onto invariant subspaces of the representation of the holonomy group
at one fixed point on the manifold.
Further solutions may exist at special values of $\a'$, 
but such sporadic solutions, although well-known \cite{rev}
in the VME, are inaccessible in the perturbative 
formulation of the unified system.

\subsection{Alternate forms of the central charge}

Using the generalized Einstein equation and the generalized VME,
the central charge (\ref{evd}) can
be written in a variety of forms,
\bs\be
c = 2 L_a{}^a + 6 \a' (2 \Phi_a \Phi^a - 
\delhp_a \Phi^a) + \oaa
\label{varca} \ee\be
= 4 L_a{}^b L_b{}^a + 2 \a' 
\left[ L_b{}^e L_d{}^f H^{bda} H_{efa} 
+3(2 \Phi_a \Phi^a - \delhp_a \Phi^a) \right] + \oaa
\label{varcc} \ee\be
= {\rm rank}(P) + 2 \a' 
\left[ L_a{}^b L_c{}^e (4L_d{}^f - 3 \delta_d{}^f ) H^{acd} H_{bef} 
+3(2 \Phi_a \Phi^a - \delhp_a \Phi^a) \right] + \oaa
\label{varcd} \ee\be
= {\rm rank}(P) + 2 \a' 
\left[  3 L^{ab} \hat{R}^+_{ab} + 
L_a{}^b L_c{}^e (4L_d{}^f - 3 \delta_d{}^f ) H^{acd} H_{bef} 
+6( \Phi_a \Phi^a - \delhp_a \Phi^a) \right] + \oaa
\label{varce} \ee\be
= 4 L_a{}^b L_b{}^a + 2 \a' 
\left[  3 L^{ab} \hat{R}^+_{ab} + 
L_a{}^b L_c{}^e  H^{acd} H_{bed} 
+6( \Phi_a \Phi^a - \delhp_a \Phi^a) \right] + \oaa
\label{varcf} \ee
\label{varc} \es
and the same forms are obtained for $\bar{c}$ by the substitution
$L\rightarrow \bar{L}$, $P\rightarrow \bar{P}$, $\Phi^a \rightarrow
\bar{\Phi}^a$ and $+\rightarrow -$.

The form of the central charge
in (\ref{varcc}) follows from (\ref{varca}) by the
generalized VME, and similarly the form in
(\ref{varcd}) follows from (\ref{varca})
by using the generalized VME in the form (\ref{ltracea}). 
The form in (\ref{varce}) then follows from (\ref{varcc})
by using the trace of the generalized Einstein equations (\ref{eva}),
(\ref{evba}),
\be
L^{ab} \hat{R}^+_{ab} - \delhp_a \Phi^a = \oa,
\qquad
\bar{L}^{ab} \hat{R}^-_{ab} - \delhm_a \bar{\Phi}^a = \oa,
\ee
and the final form in (\ref{varcf}) follows from (\ref{varce})
by another application of the generalized VME in the form (\ref{ltraceb}).

The first form in (\ref{varca}) is the `affine-Virasoro form' of
the central charge, noted above as possibly exact to all orders 
in the general sigma model. 
The form in (\ref{varce}), with the first occurence
of the generalized Ricci tensor, will be called the `conventional form'
of the central charge because it reduces easily to the central charge
of the conventional stress tensors
\be
c_G(\Phi) =\dim (M) + 3 \a' (4 | \del \Phi |^2 - 4 \del^2 \Phi + R + 
\frac{1}{12} H^2) + \oaa
\ee
when $P=G$ and $\Phi_a^G=\del_a \Phi$. The conventional form is also
the form in which
we found it most convenient
to prove the constancy of $c$ and $\bar{c}$ (see Section~5.7). 
The final form of $c$ in (\ref{varcf}) is the form which we believe
comes out directly from the two-loop computation (see Section~6).

\subsection{Bianchi identities and constant central charge}

In this section, we check that the rest of the unified system
implies that the central charges are constant
\be
\dif_i c = \delhp_i c = \oaa, \qquad \dif_i \bar{c} = 
\delhm_i \bar{c} = \oaa
\ee
although we give details only for $c$. 

For this check, we use the zeroth-order form of the generalized VME and
Einstein equations
\bs\be
P_a{}^c P_c{}^b = P_a{}^b, \qquad \delhp_i P_a{}^b=0 
\ee\be
\beta_{ab} \equiv P^{cd} \hat{R}^+_{acdb} +
2 \delhp_a \Phi^{(0)}_b =0
\ee\be
\Phi^{(0)}_a=P_a{}^b \Phi_b^{(0)}
\label{auxx9b}
\ee
\be
\hat{R}^+_{cd}{}^{ae} P_e{}^b + 
\hat{R}^+_{cd}{}^{be} P_e{}^a  =0
\label{auxx9a}
\ee
\label{auxx9} \es
where (\ref{auxx9a}) is the integrability condition discussed in Section~5.5.
We will also use the $(+)$ form of the generalized Bianchi identities
\be
\eqalign{
\delh^{\pm}_a \hat{R}^{\pm}_{bcde} +& 
\delh^{\pm}_b \hat{R}^{\pm}_{cade}+
\delh^{\pm}_c \hat{R}^{\pm}_{abde} \cr
&  \pm(
H_{ab}{}^f \hat{R}^{\pm}_{fcde} +
H_{bc}{}^f \hat{R}^{\pm}_{fade}+
H_{ca}{}^f \hat{R}^{\pm}_{fbde} ) =0 \cr}
\label{bianchi}
\ee
which follow from the explicit form of the generalized Riemann tensors
in (\ref{tordeff}).

The form of the central charge which we will use is the `conventional
form' in (\ref{varce}),
\bs
\vskip -.50in
\bea
\!\!\! c &\!\!\!\!\! =  & \!\!\!\!\!{\rm rank}(P) + 6 \a' \beta_{\Phi} + \oaa
\\{} 
\!\!\! \beta_{\Phi} & \!\!\!\!\! \equiv &\!\!\!\!\! 
-\frac{1}{2} P^{ab} \hat{R}^+_{cab}{}^c + 
P_a{}^b P_c{}^e (\frac{1}{6} P_d{}^f - \frac{1}{4}
 \delta_d{}^f ) H^{acd} H_{bef} 
+2( \Phi^{(0)}_a \Phi_{(0)}^a - \delhp_a \Phi_{(0)}^a) 
\eea
\es
so that the check is equivalent to showing that $\beta_{\Phi}$
is a constant. As we shall see, the desired relation emerges in the
form
\be
0= -\delhp_b \beta_a{}^b + \beta^b{}_c H^c{}_{ba} +2
\beta_a{}^b \Phi^{(0)}_b = \delhp_a \beta_{\Phi} 
\label{aux111}
\ee
where $\beta_{ab}=0$ is the generalized Einstein equation.

We outline the steps needed to verify (\ref{aux111}).
For the first term, we find
\bs\bea
\delhp_b \beta_{a}{}^b & = & 
\delhp_b (P^{cd} \hat{R}^+_{acd}{}^b) + 
2 \delhp_b \delhp_a \Phi^b_{(0)}
\label{aux12a} \\{}
& = & \delhp_b (P^{cd} \hat{R}^+_{acd}{}^b) + 
2 [\delhp_b ,\delhp_a] \Phi^b_{(0)}
+2 \delhp_a \delhp_b \Phi^b_{(0)}
\label{aux12b} \\{}
 & =&   \delhp_a(  \frac{1}{2} P^{bc} \hat{R}^+_{dbc}{}^d
 + 2 \delhp_b \Phi^b_{(0)} ) +
 P^{bc} (H_{af}{}^d \hat{R}^+_{dbc}{}^f - \frac{1}{2} H_{cf}{}^d 
\hat{R}^+_{dab}{}^f) \nonu
&  & + 2 \hat{R}^+_{acb}{}^c \Phi^b_{(0)} + 
 2 H_{ab}{}^c \del_c \Phi^b_{(0)} 
\label{aux12c} \eea \label{aux12} \es
where we used $P$ times the $(+)$ Bianchi identity (\ref{bianchi})
and the relation (\ref{tordefe}) to obtain (\ref{aux12c}). For the
dilaton vector term we find
\bs\be
2 \beta_a{}^b \Phi^{(0)}_b = 
2 \delhp_a (\Phi^{(0)}_b \Phi_{(0)}^b) - 
2 \hat{R}^+_{ac}{}^{bd} P_d{}^c \Phi_b^{(0)}
\label{aux14a}
\ee\be =
2 \delhp_a (\Phi^{(0)}_b \Phi_{(0)}^b) +
2 \hat{R}^+_{ac}{}^{cb}  \Phi_b^{(0)}
\label{aux14b}
\ee
\label{aux14} \es
where we used (\ref{auxx9b}) and (\ref{auxx9a}) to obtain (\ref{aux14b}).
Adding all three terms of (\ref{aux111}), we find that
\bs\be
0= -\delhp_b \beta_a{}^b + \beta^b{}_c H^c{}_{ba} +2
\beta_a{}^b \Phi^{(0)}_b 
 \qquad \qquad \qquad \qquad \qquad \qquad
\label{aux15a}
\ee\be
=   \delhp_a (- \frac{1}{2} P^{bc} \hat{R}^+_{dbc}{}^d
 - 2 \delhp_b \Phi^b_{(0)} + 2 \delhp_b \Phi^b_{(0)} ) 
 - \frac{1}{2} \hat{R}^+_{ade}{}^b P_b{}^c H_c{}^{de}
\label{aux15b}
\ee\be
= 
\delhp_a \beta_{\Phi} \label{aux15c}
\qquad \qquad \qquad \qquad \qquad \qquad 
 \qquad \qquad \qquad \qquad \qquad
\ee \label{aux15} \es
where the identity (\ref{dtracea}) was used to write the last term in
 (\ref{aux15b}) as a total derivative. This completes the
demonstration that the central charge $c$ is a constant, and
similar steps using the $(-)$ Bianchi identity in (\ref{bianchi})
show that $\bar{c}$ is also constant.

\subsection{Solution classes and a simplification}

\vs .4cm
\noindent \underline{Class I and Class II solutions}
\vs .3cm 

The solutions of the unified system (\ref{ev}) can be divided into
two classes,

\vs .1cm

\noindent
Class I. $T$ conformal but $T_G(\Phi_a)$ not conformal:
\bs\be
P^{cd} \hat{R}^+_{acdb} +  2 \delhp_a \Phi_b^{(0)} = 0,
\qquad P_a{}^b \Phi_b^{(0)} = \Phi_a^{(0)}
\ee\be
 \hat{R}^+_{ab} -2  \delhp_a  \Phi_b^{(0)G}  \neq 0
\ee\es

\vs .1cm

\noindent
Class II. $T$ and $T_G(\Phi_a)$ both conformal:
\bs\be
P^{cd} \hat{R}^+_{acdb} + 2 \delhp_a \Phi_b^{(0)} = 0,
\qquad P_a{}^b \Phi_b^{(0)} = \Phi_a^{(0)}
\label{pss}
\ee\be
 \hat{R}^+_{ab} -2  \delhp_a  \Phi_b^{(0)G}  = 0
\label{auxn14}
\ee\es

\vs .1cm

\noindent
where $\Phi_b^{(0)G}$ is unrestricted. The 
distinction here is based on whether or not (in addition
to the generalized Einstein equations) the dilaton-vector Einstein
equations in (\ref{qqa}) 
are {\it also} satisfied. In the case when
the dilaton solution $\Phi_a(\Phi)$ in
(\ref{eig}) is taken for the dilaton vector,
the question is whether or not the background sigma model is itself
conformal in the conventional sense.

In Class I, we are constructing a conformal stress tensor $T$ in the
operator algebra of a sigma model whose conventional stress tensor 
$T_G(\Phi_a)$ with general dilaton vector is not conformal.
This is a situation not encountered in the general
affine-Virasoro construction because the conventional stress-tensor 
$T_g$ of the WZW model is the affine-Sugawara construction,
which is conformal. It is expected that Class I solutions are generic
in the unified system, since there are so many non-conformal sigma
models, but there are so far no non-trivial\footnote{Trivial
examples in Class I
are easily constructed as tensor products of conformal and
non-conformal theories (see also Section~5.5).}
examples (see however \cite{chun},
which proposes a large set of candidates).

In Class II, we are constructing a conformal stress tensor $T$ in
the operator algebra of a sigma model whose conventional stress
tensor $T_G(\Phi_a)$ with general dilaton vector
is conformal. This class includes the 
case where the conventional stress tensors $T_G(\Phi)$ are conformal
so that the sigma model is conformal in the conventional sense.
The general affine-Virasoro
construction provides a large set of non-trivial examples in Class II
when the sigma model is the WZW action.
Other examples are known from the general affine-Virasoro construction
which are based on coset constructions, 
instead of WZW. In particular, \cite{lie}
 constructs exact Virasoro
operators in the Hilbert space of a certain class of
$g/h$ coset constructions,
and it would be interesting to study these operators
as Class II solutions in the
sigma model description of the coset constructions.

It is also useful to subdivide Class II solutions into Class IIa and IIb.
In Class IIb, we require the natural identification
\be
\Phi_a = 2 L_a{}^b \Phi_b^G + \oa
\label{aabb}
\ee
which solves (\ref{evn}), and Class IIa is the set of solutions in Class II
without this identification. Note in particular that Class IIb contains
all solutions in Class II with the dilaton solution $\Phi_a(\Phi)$ in
(\ref{eig}). We also remark that, for solutions in
Class IIb, the Ricci form (\ref{auxn12})
 of the generalized Einstein equations
can be written as
\be
(\hat{R}^+_{ab} - 2 \delhp_a \Phi_b^G ) L^{bc} = \oa
\label{abcd}
\ee
where the left factor is the dilaton-vector Einstein equation in
(\ref{qqc}).

\vs .4cm
\noindent \underline{Simplification for Class IIb}
\vs .3cm 

For solutions in Class IIb, the unified system (\ref{ev}),
(\ref{evball}) can be written in a simpler form, where the
generalized Einstein equations (\ref{eva}) and (\ref{evba}) are 
replaced by the dilaton-vector Einstein equations 
\be
\hat{R}^+_{ab} - 2 \delhp_a \Phi^G_b = \oa, \qquad
\hat{R}^-_{ab} - 2 \delhm_a \bar{\Phi}^G_b = \oa.
\label{z3}
\ee
This simplification
of the system follows for solutions in Class IIb because the Ricci
form (\ref{abcd}) of the generalized Einstein equations are automatically
solved when the dilaton-vector Einstein equations are satisfied (and
the Ricci form is equivalent to the generalized Einstein equations
when the covariant-constancy conditions (\ref{xyzc}) are
satisfied).

A further simplification in Class IIb follows for the dilaton solution
$\Phi_a(\Phi)$. In this case, equations (\ref{z3}) can be replaced by
the conventional Einstein equations, so that the unified system reads
\bs\be
R_{ij}+\frac{1}{4} (H^2)_{ij} - 2 \del_i \del_j \Phi = 
{\cal O}(\a ') 
\label{xyza}
\ee\be
\del^k H_{kij} - 2 \del^k \Phi H_{kij} = {\cal O}(\a')
\label{xyzb}
\ee\be
\delhp_i L^{ab} = {\cal O}(\a') , \qquad
\delhm_i \bar{L}^{ab} = {\cal O}(\a') 
\label{xyzc}
\ee\be
\eqalign{ L^{ab}  = & 2 L^{ac} G_{cd} L^{db} \cr
 & - \a' (L^{cd} L^{ef} H_{ce}{}^a H_{df}{}^b + 
 L^{cd} H_{ce}{}^f H_{df}^{(a} L^{b)e} ) \cr
& - \a' (L^{c(a} G^{b)d} \del_{[c} \Phi_{d]} ) + {\cal O}(\a'^2) \cr}
\label{xyzd}
\ee\be
\eqalign{ \bar{L}^{ab}  = & 2 \bar{L}^{ac} G_{cd} \bar{L}^{db} \cr
 & - \a' (\bar{L}^{cd} \bar{L}^{ef} H_{ce}{}^a H_{df}{}^b + 
 \bar{L}^{cd} H_{ce}{}^f H_{df}^{(a} \bar{L}^{b)e} ) \cr
& - \a' (\bar{L}^{c(a} G^{b)d} 
\del_{[c} \bar{\Phi}_{d]} ) + {\cal O}(\a'^2) \cr}
\label{xyze}
\ee\be
c = 2 G_{ab} L^{ab} + 6 \a' (2 \Phi_a \Phi^a - \del_a \Phi^a)
 + {\cal O}(\a'^2)
\label{xyzf}
\ee\be
\bar{c} = 2 G_{ab} \bar{L}^{ab} + 6 \a' (2 \bar{\Phi}_a \bar{\Phi}^a 
- \del_a \bar{\Phi}^a)
 + {\cal O}(\a'^2).
\label{xyzg}
\ee\be
\Phi_a=\Phi_a(\Phi)=2 L_a{}^b \del_b \Phi, \qquad 
\bar{\Phi}_a=
\bar{\Phi}_a(\Phi) = 
2\bar{L}_a{}^b \del_b \Phi.
\label{xyzh}
\ee
\label{xyz} \es
This form of the system is close in spirit to the VME of the
general affine-Virasoro construction: The solution of the conventional
Einstein equations in (\ref{xyza}), (\ref{xyzb}) provides a 
conformal background, in which we need
only look for solutions of the generalized VMEs in the form
\be
L_a{}^{b} = \frac{P_a{}^{b}}{2} + \oa, \qquad
\bar{L}_a{}^{b} = \frac{\bar{P}_a{}^{b}}{2} + \oa
\ee
where $P$ and $\bar{P}$ are covariantly constant projectors.
Moreover, as in the VME, we shall see below that all solutions of the
simplified system (\ref{xyz})
exhibit $K$-conjugation covariance.

\subsection{$K$-conjugation covariance}

For Class II solutions, both the generalized and the 
dilaton-vector Einstein equations are satisfied,
\bs\be
L^{cd} \hat{R}^+_{acdb} + \delhp_a \Phi_b = \oa 
, \qquad \Phi_a = 2 L_a{}^{b}  \Phi_b + \oa
\label{z1}
\ee\be
L_G^{cd} \hat{R}^+_{acdb} + \delhp_a \Phi^G_b = \oa ,
\ee\es
so that both the generalized and conventional stress tensors
\bs\be
T = - L^{ab} (\frac{\Pi_a \Pi_b}{\a '} + \frac{1}{2} \Pi_c \Pi_d 
H_{ae}{}^c H_b{}^{ed} ) + \dif (\Pi_a \Phi^a) + {\cal O}(\a')
\ee\be
T_G(\Phi_a) = - L_G^{ab} (\frac{\Pi_a \Pi_b}{\a '} + \frac{1}{2} \Pi_c \Pi_d 
H_{ae}{}^c H_b{}^{ed} ) + \dif (\Pi_a \Phi_G^a) + {\cal O}(\a')
\ee \label{auxn15} \es
are conformal. The explicit form of $L_G$ is given in (\ref{deflg}).

Experience with the general affine-Virasoro construction leads us
to suspect that Class II solutions of the unified system will exhibit
$K$-conjugation covariance (see Section~2.1), which is the statement
that the $K$-conjugate stress tensor $\tilde{T}$,
\be
\tilde{T}\equiv T_G-T, \qquad \tilde{c}=c_G-c
\label{covprop}
\ee
is also conformal when $T_G$ and $T$ are conformal. 
Together, the conformal stress tensors $T$ and $\tilde{T}$ are
called a $K$-conjugate pair, and the $K$-conjugate pairs
$T=T_G$, $\tilde{T}=0$ are a set of trivial examples in Class II.
In what follows,
we will demonstrate that (through one loop) $K$-conjugation
covariance is indeed a property of a large subset
of the solutions in Class II.
A parallel demonstration can be
given for the antiholomorphic sector, which is not discussed
explicitly here. 

We begin by translating the algebraic
covariance (\ref{covprop}) into the geometric
system by defining the K-conjugate inverse inertia tensor $\tilde{L}^{ab}$,
\bs\be
\tT = -\tL^{ab} (\frac{\Pi_a\Pi_b}{\a'} + \frac{1}{2} \Pi_c \Pi_d H_{ae}{}^c
 H_b{}^{ed}) + \dif(\Pi_a \tP^a) + \oaa
\label{tconj}
\ee\be
\tc=2G_{ab} \tL^{ab} + 6\a' (2 \tP^a \tP_a - \delhp_a \tP^a) + \oaa
\label{cconj}
\ee
\vskip -.15in
\be
\tL^{ab} = L^{ab}_G - L^{ab}, \qquad \tP_a=\Phi_a^G-\Phi^a
\label{lconj}
\ee \label{conj} \es
where we have used (\ref{auxn15})
 and the `affine-Virasoro form' (\ref{varca}) of
the central charge to obtain (\ref{cconj}), (\ref{lconj}). 
The eigenvalue relation for the $K$-conjugate dilaton vector
$\tP_a$,
\be
2 \tL_a{}^b \tP_b = \tP_a + \oa
\ee
follows from (\ref{z1}) and (\ref{lconj}). This relation is necessary for
conformal invariance of the $K$-conjugate theory, and, taken together with
(\ref{pss}) and (\ref{lconj}), also implies that
\be
\Phi_a = 2 L_a{}^b \Phi_b^G + \oa , \qquad
\tP_a = 2 \tL_a{}^b \Phi_b^G + \oa.
\label{zz3}
\ee
These relations show that $K$-conjugation covariance is restricted to solutions
in Class IIb (see (\ref{aabb})).

To verify $K$-conjugation covariance, 
we must then show that the conjugate
relations
\bs\be
\tL^{cd} \hat{R}^+_{acdb} + \delhp_a \tP_b = \oa 
\qquad \qquad \qquad \qquad
\label{conjeva}
\ee\be
\delhp_i \tL^{ab} = \oa 
\qquad \qquad \qquad \qquad
\label{conjevb}
\ee\be
\eqalign{\tL^{ab} = & 2 \tL^{ac} G_{cd} \tL^{db} \cr
& - \a' (\tL^{cd} \tL^{ef} H_{ce}{}^a H_{df}{}^b + \tL^{cd} H_{ce}{}^f
 H_{df}^{(a}\tL^{b)e} ) \cr
& -\a' \tL^{c(a} G^{b)d} \del_{[c} \tP_{d]} \cr}
\label{conjevc}
\ee \label{conjev} \es
hold when the same equations are satisfied by $L^{ab}$ and $L_G^{ab}$.
Because the generalized Einstein equations and covariant-constancy condition
are linear in $L$, the relations (\ref{conjeva}), (\ref{conjevb}) follow
immediately from the corresponding relations for $L$ and $L_G$.

To verify the $K$-conjugate VME (\ref{conjevc}), we need the semiclassical
expansion introduced in Section~5.4,
\bs\be
L=2L^2 + \a'(X_L + X_{\Phi}) + \oaa
\label{semia} \ee\be
L=\frac{1}{2} P + \a' (\Delta_L + \Delta_{\Phi} ) + \oaa
\label{semib} \ee\be
P^2=P, \qquad \Delta_L=\Delta_L P + P \Delta_L + X_L
\label{semic} \ee\be
\Delta_{\Phi}= \Delta_{\Phi} P + 
P \Delta_{\Phi}  + X_{\Phi}
\label{semid} \ee \label{semi} \es
where $(AB)_a{}^b=A_a{}^c B_c{}^b$. Here we have divided the quantities
$X=X_L+X_{\Phi}$ and $\Delta=\Delta_L+\Delta_{\Phi}$ of Section~5.4 into
the ordinary ($L$) contributions and the dilatonic ($\Phi$) contributions,
e.g.
\be X_{\Phi}=\frac{1}{2} (FP-PF), \qquad F_{ab}=
\del_{[a} \Phi^{(0)}_{b]}
\ee
where $F$ is the zeroth-order form of the field strength of the dilaton
vector.

We also need the corresponding expansion of the K-conjugate quantities
\bs\be
\tL = \frac{1}{2} \tilde{P} + \a(\tilde{\Delta}_L + 
\tilde{\Delta}_{\Phi} ) + \oaa
\label{aux17a}
\ee\be
\tilde{P}=1-P, \qquad \tilde{\Delta}_L= -\frac{1}{4} H^2 - \Delta_L,
\qquad \tilde{\Delta}_{\Phi}= - \Delta_{\Phi}
\label{aux17b}
\ee \be
\tilde{F} = F_G - F = \del_{[a} \tilde{\Phi}_{b]}^{(0)}
\label{aux17c}
\ee
\label{aux17} \es
where the relations in (\ref{aux17b}), (\ref{aux17c})
 were obtained from (\ref{lconj})
and the explicit form of $L_G$ in (\ref{deflg}). The absence of a 
$G$-term in $\tilde{\Delta}_{\Phi}$ follows because the dilaton 
vector fails to contribute to $L_G$,
\be
\Delta_{\Phi}^G = X_{\Phi}^G =  0
\label{aux18a} 
\ee
as seen in Section~5.1.

$K$-conjugation covariance for the ``ordinary terms'' of the generalized
VME is established by verifying the identity
\bs\be
\tilde{\Delta}_L \tilde{P} + \tilde{P}  \tilde{\Delta}_L 
+\tilde{X}_L = \tilde{\Delta}_L
\ee\be
\tilde{X}_L  = X_G-X_L = X_L(\tilde{P}/2)
\ee\es
using (\ref{aux17b}) and the generalized VME in the form (\ref{semic}).

To study $K$-conjugation for the dilatonic contribution, we evaluate
the quantity
\bs\be
\tilde{\Delta}_{\Phi} \tilde{P} +
\tilde{P} \tilde{\Delta}_{\Phi}  + \frac{1}{2}
(\tilde{F} \tilde{P} - \tilde{P} \tilde{F}) 
\label{aux19a} \ee\be
=\tilde{\Delta}_{\Phi} + \frac{1}{2} (PF_G-F_GP)
\label{aux19b}
\ee \label{aux19} \es
using (\ref{aux17b}), (\ref{aux17c}) and the generalized VME in the
form (\ref{semid}). When
\be
PF_G-F_GP = P^{c(a} G^{b)d} F^G_{cd} = 0
\label{z2}
\ee
holds, the relation in (\ref{aux19}) is the $K$-conjugate form of the linear
equation (\ref{semid}) for the dilatonic contribution to $L$. It follows
that the condition (\ref{z2}) is necessary and sufficient for $K$-conjugation
covariance of the dilatonic contribution. 

One solution of the condition (\ref{z2}) is 
\be F_{ab}^G=0 \label{zz2} \ee
which implies the dilaton solution $\Phi^{(0)G}_a = \del_a \Phi$
for $T_G$. In this case, the conditions (\ref{zz3}) imply that the dilaton
vectors are given by the dilaton solution $\Phi_a(\Phi)$, uniformly
for $T$, $\tilde{T}$ and $T_G$, and the unified system reduces to the
simplified system (\ref{xyz}).

In summary, we find that all solutions of the simplified system (\ref{xyz})
exhibit $K$-conjugation covariance at one loop. In particular, this
includes all solutions with zero dilaton vector. 
To find other $K$-conjugate pairs in Class IIb,
one must find other solutions of (\ref{z2}).

We finally comment on some open questions in this direction. In the 
first place, it should be checked that the K-conjugate
pair of stress tensors $T$ and $\tT$ commute at one-loop level,
as expected from the general affine-Virasoro construction.
Second, we remark that the discussion above does not in fact contain
any explicit examples of $K$-conjugation covariance for the VME
with non-trivial improvement: 
In this case, the conditions (\ref{z2}) and (\ref{zz2}) read
respectively
\bs\be
f_{cd}{}^{(a} L^{b)c} D^d_g = {\cal O}(k^{-3}) 
\label{dc1}
\ee\be
f_{ab}{}^c D_c^g = {\cal O}(k^{-2})
\label{dc2}
\ee\es
and, to the order at which we are working,
 the solution of (\ref{dc2}) is $D_g= 0$. Using 
the relations (\ref{consb}) and (\ref{zz3}), this then
implies that $D=\tilde{D}=0$.
Therefore, 
the only possibility to find $K$-conjugation covariance in the
VME with non-trivial improvement is to find solutions of (\ref{dc1})
with $D_g \neq 0$. 


\section{The computation}

In this section, we outline the one-loop calculation which gives the
unified Einstein-Virasoro master equations
 (\ref{ev}) and (\ref{evball}), giving details only 
for the holomorphic system. The corresponding results for the 
antiholomorphic system follow from these results by the rule
noted in Section~4. 

The starting point of the computation is
\bs\be
S=\frac{1}{2\a'} 
\int d^2 z \P^a \BP^b (G_{ab}+B_{ab}) 
\label{aux20a}
\ee\be
T=-\frac{1}{2\a'} \l_{ab} \Pi^a \Pi^b + \dif(\Pi^a \Phi_a) + 
(\mbox{{\rm one loop counterterms}}) + \oa
\label{aux20b}
\ee
\be
\l_a{}^b =\l_a{}^c \l_c{}^b + \oa, \qquad
\delhp_c \l_a{}^b = \oa \label{aux20d}
\ee\be
<T(z)T(w)> = \frac{c/2}{(z-w)^4} + 2 \frac{<T(w)>}{(z-w)^2} +
 \frac{<\dif T(w)>}{(z-w)} + {\rm reg.}
\label{aux20e}
\ee \label{aux20} \es
where $\Pi,\BP$ are defined in (\ref{defpi}) and we have introduced the 
scaled
inverse inertia tensor $\l_a{}^b \equiv 2 L_a{}^b$. For convenience,
we have also included the classical results (\ref{aux20d}) 
of Section~3 in the starting point, though these equations can also
be obtained from the tree graphs in the computation below. 


By including the dilaton vector
in the stress tensor instead of the action,
we are following Banks, Nemeschansky and Sen \cite{sig12}, who showed that this 
language correctly reproduces the beta functions \cite{sig9,sig12}
 of the sigma model
when $T \rightarrow T_G(\Phi)$
 is the conventional stress tensor (\ref{zz1})
of the
sigma model. Although the stress tensor method is computationally more
involved than the usual action method, it is clearly the correct language
in which to study the general Virasoro construction in the general
sigma model. 

\subsection{Background field expansion}

A central idea in the method of Ref. \cite{sig12} is to use a covariant 
background field expansion to evaluate 
the left and right sides of (\ref{aux20e}) in perturbation theory.
An efficient algorithm for this expansion
has been provided by Mukhi \cite{sig10}, 
generalizing earlier work in \cite{sig1a,sig2,sig4}.

One way to write down this expansion is
to introduce an operator $M$ which satisfies
\bs\be
M(T_{a_1, \ldots, a_n}(x)) = y^a \del_a T_{a_1,\ldots, a_n}(x)
\label{aux21a} 
\ee\be
M(\Pi^a) = \del y^a, \qquad M(\BP^a)=\delb y^a 
\label{aux21b} \ee\be
M(\del y^a) = -R^a{}_{bcd}(x) y^b y^c \Pi^d , \qquad
M(\delb y^a) = -R^a{}_{bcd}(x) y^b y^c \BP^d ,
\label{aux21c} \ee\be
\del A^a(x) = \dif A^a(x) + \Pi^b A^c(x) \omega_{bc}{}^a, \qquad
\delb A^a(x) = \dbar A^a(x) + \BP^b A^c(x) \omega_{bc}{}^a
\label{aux21d} \ee\be
\Pi^a=e_i{}^a \dif x^i , \qquad \BP^a = e_i{}^a \dbar x^i
\label{aux21e} \ee \label{aux21} \es
where $x^i$, $i=1,\ldots,\dim(M)$ are the background fields and
$y^a$, $a=1,\ldots,\dim(M)$ are the quantum fields. The background
fields $\Pi$ and $\Pb$ in (\ref{aux21e})
satisfy the classical equations of motion (\ref{eqmot}),
which we use here in the form
\be \del \BP^a = \delb \Pi^a = -\frac{1}{2} \P^c \BP^d H_{cd}{}^a
\label{eqmot2}
\ee
where $\del$ and $\delb$ are defined in (\ref{aux21d}). Then, one
uses 
the operator $M$ to expand any scalar function $\Psi$
(such as $S$ or $\Pi_a \Phi^a$) in the form
\be
\Psi=\sum_{n=0}^{\infty} \Psi^{(n)}, 
\qquad \Psi^{(n)}= \frac{1}{n} M(\Psi^{(n-1)})
\label{j2}
\ee
where $n$ is the number of quantum fields $y$.

Our goal is to evaluate the Virasoro condition
(\ref{aux20e}) through one loop. For this it
is sufficient to compute only the fourth and second order pole terms
in $<T(z)T(w)>$ (because this and the relation $<T(z)T(w)>=<T(w)T(z)>$,
which holds in Euclidean perturbation theory, then
fix the remaining singular terms). 
Since the background
fields $\Pi,\bar{\Pi}$ have dimension one, we need 
only look at the contributions to $<T(z)T(w)>$ which involve
at most two background fields. Now consider an arbitrary one-loop Feynman
diagram in the expansion of $<T(z)T(w)>$. The diagram contains a
certain number $n(b,q)$ of 
vertices with $b$ background fields and $q$ quantum fields, 
so that
\be
l=1 + \frac{1}{2} \sum_{b,q \geq 0} n(b,q) (q-2).
\label{j1}
\ee
Ignoring the dilatonic terms in $T$ for the moment,
we see from the form
of $S$ in (\ref{aux20a}) and $T$ in (\ref{aux20b}) that $b+q \geq 2$. 
If we denote the total number of background fields in the diagram
by $b_{\rm tot}\equiv \sum_{b,q} bn(b,q)$, we find from
(\ref{j1}) that
\be b_{\rm tot} + 2l -2 = \sum_{b,q \geq 0} n(b,q) (b+q-2).
\ee
Each of the terms on the right hand side is nonnegative, and 
if $n(b,q)\neq 0$ this implies that
\be b+q \leq b_{\rm tot} + 2l.
\label{j3}
\ee
In our case we have $l=1$ and $b_{\rm tot} \leq 2$, so that we need
only those terms with $b+q\leq 4$ from the
background field expansion of $S$ and $T$. For the terms involving
the dilaton vector, which involve an extra factor of $\a'$, we find in a
similar fashion that we need only
the terms with $b+q\leq 2$. Notice that a derivative of a background
field, such as $\del \P^a$, has conformal weight two and therefore counts
as $b=2$.

Using the iteration
(\ref{j2}), we find for these terms the following results
\bs\be
S^{(0)}  =  \frac{1}{2\a'} \int d^2 z \P^a\Pb^b(G_{ab} + B_{ab})
\ee\be
S^{(1)}  =  \frac{1}{2\a'} \int d^2 z (\P_a \deb y^a +
\Pb_a \de y^a + y^c H_{cab} \P^a \Pb^b) 
\ee\be
\eqalign{
S^{(2)}  =  \frac{1}{2\a'} \int d^2 z ( & \de y_a \deb y^a -
 \Pb^d  y^b y^c \Pi^a R_{dbca} +
 \half y^c y^d \de_d H_{cab} \P^a \Pb^b  \cr
&  +\half y^c H_{cab} \de y^a \Pb^b
    +\half y^c H_{cab} \P^a \deb y^b ) \cr}
\ee\be
\eqalign{
S^{(3)}  =  \frac{1}{2\a'} \int d^2 z ( & 
 -\frac{2}{3} \deb y^d y^c y^b \P^a R_{dcba}
 -\frac{2}{3} \del y^d y^c y^b \Pb^a R_{dcba} \cr
 &
 + \frac{1}{3} y^c y^d \de_d H_{cab} \de y^a \Pb^b
 + \frac{1}{3} y^c y^d \de_d H_{cab} \P^a \deb y^b
 + \frac{1}{3} y^c H_{cab} \de y^a \deb y^b) \cr}
\ee\be
S^{(4)}  =  \frac{1}{2\a'} \int d^2 z (
 -\frac{1}{3} \delb y^d y^c y^b \de y^a R_{dcba}
 + \frac{1}{4} y^c y^d \de_d H_{cab} \de y^a \deb y^b )
\ee \label{sexpansion} \es
and 

\bs\be
T^{(0)}  =  -\frac{1}{2\a'} \lambda_{ab} \P^a \P^b + \de \P^a \Phi_a +
 \P^a \P^b \de_b \Phi_a 
\ee\be
T^{(1)}  =  - \frac{1}{\a'}
\lambda_{ab} \P^a \de y^b -\frac{1}{2\a'} \P^a \P^b y^c
 \de_c \lambda_{ab} + (\de \de y^a) \Phi_a  
+  (\de y^a \P^b + \de y^b \P^a) \de_b \Phi_a 
\ee\be
\eqalign{
T^{(2)}  = &  -\frac{1}{2\a'}
 \lambda_{ab} \de y^a \de y^b - \frac{1}{\a'} \P^a \de y^b y^c
\de_c \lambda_{ab} -\frac{1}{4\a'} \P^a \P^b y^c y^d \de_d \de_c \lambda_{ab} 
\cr & 
 + \frac{1}{2\a'} \P^d y^b y^c R_{dbc}{}^e \lambda_{ea} \P^a \cr}
\ee\be
\eqalign{
T^{(3)}  = & -\frac{1}{2\a'}  \P^a \de y^b y^c y^d \de_d \de_c \lambda_{ab}
 - \frac{1}{2\a'} \de y^a \de y^b y^c \de_c \lambda_{ab} \cr & 
 +\frac{1}{6\a'} \del y^d y^b y^c R_{dbc}{}^e \lambda_{ea} \P^a + 
 \frac{1}{2\a'} \P^d y^b y^c R_{dbc}{}^e \lambda_{ea} \del y^a \cr }
\ee
\label{texpansion}
\es

\noindent
where the superscripts in $S$ and $T$ indicate 
the number of quantum fields in the expansion. The first-order term
$S^{(1)}$ of the action is zero by the classical equations of motion,
and we have omitted terms with $b+q=3$, such as
$\del \P^a y^b \del_b \Phi_a$ and $\Pi^a y^b \Pi^c \del_c \del_b \Phi_a$
in $T^{(1)}$. We have also dropped a term proportional to 
$\del^2 (y^a y^b) \del_a \Phi_b$ in $T^{(2)}$, and
omitted $T^{(4)}$. Although such terms should in principle be included
according to our previous discussion, an inspection of the table of
possible Feynman diagrams in Figure~1 shows that they never appear. 

\begin{figure}
\epsfig{figure=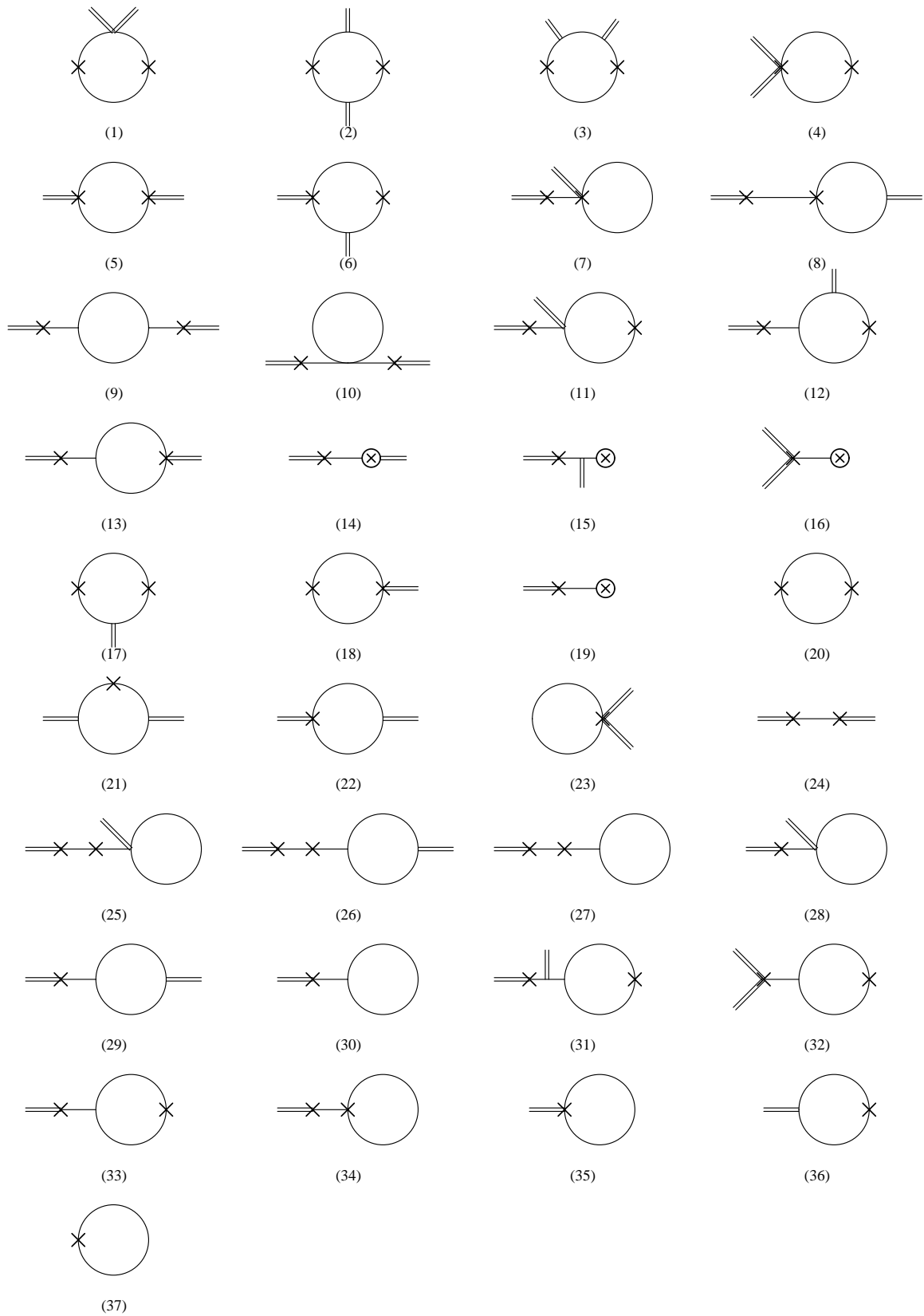,height=8.6in,width=6.0in}
\caption{A table of Feynman diagrams; crosses ($\times$)
indicate vertices from
the non-dilaton part of $T$, encircled crosses ($\otimes$) 
indicate vertices coming from the dilaton part of $T$, double lines
indicate background fields and single lines indicate quantum fields.}
\end{figure}

The background field expansions of the action $S$ and the stress tensor
$T$ contain many terms that depend on the spin connections. Some
of them appear only in tensors like $R_{abcd}$ and $\del_a H_{bcd}$
that are covariant with respect to target-space 
local Lorentz transformations.
In Feynman diagrams, such terms yield only covariant contributions.
However, there are also spin connections in $\del y^a$ and $\deb y^a$,
and these give non-covariant contributions to Feynman diagrams
involving the corresponding vertices. Since both $S$ and $T$ are
Lorentz invariant, our calculation should yield a Lorentz invariant answer.
The only possible non-covariant contributions come from the diagrams
involving the spin connection in $\del y^a$ and $\deb y^a$, and 
these spin connections appear algebraically, i.e. without a derivative.
There are no purely algebraic Lorentz invariant tensors one can
build out of the spin-connection, and therefore,
unless there are sigma model 
anomalies \cite{moore}, 
 all diagrams
involving spin connections from $\del y^a$ and $\deb y^a$ should
cancel. Sigma model
anomalies are usually caused by chiral fermions, which are absent in
our case, so it would seem correct to 
simplify the algebra by setting all spin connections
in $\del y^a$ and $\deb y^a$ 
equal to zero. However, we have decided to check these statements by
including all spin connections in our computations,
and we will see that they do indeed cancel.

To proceed, it will be convenient to rearrange these expansions
by eliminating $\del$ in favor of $\dif$ using (\ref{aux21d}). At the
same time, we will use (\ref{aux20d}) to eliminate all
derivatives of $\lambda_{ab}$, and contractions between $\lambda$'s.
When we do this in the expression obtained from a one-loop
diagram, this induces errors only at two loops. The only case where
we cannot do this is for the tree diagram (24) in Figure~1, where
it would induce errors at one-loop, and we have to treat diagram (24)
with a little more care. 

After this rearrangement, the expansions
(\ref{sexpansion}) and (\ref{texpansion}) become
\bs\be
\eqalign{
S^{(2)}  =  \frac{1}{2\a'} \int d^2 z ( & \dif y_a \dbar y^a -
 \Pb^d  y^b y^c \Pi^a R_{dbca} +
 \half y^c y^d Z^1_{dcab} \P^a \Pb^b  \cr
&  + y^c \hop_{bca} \dif y^a \Pb^b
    - y^c \hom_{abc} \P^a \dbar y^b ) \cr}
\label{sexpansion2a}
\ee\be
\eqalign{
S^{(3)}  =  \frac{1}{2\a'} \int d^2 z ( & 
 -\frac{2}{3} \dbar y^d y^c y^b \P^a R_{dcba}
 -\frac{2}{3} \dif y^d y^c y^b \Pb^a R_{dcba} \cr
 &
 + \frac{1}{3} y^c y^d Z^2_{dcab} \dif y^a \Pb^b
 + \frac{1}{3} y^c y^d Z^3_{dcab} \P^a \dbar y^b
 + \frac{1}{3} y^c H_{cab} \dif y^a \dbar y^b) \cr}
\ee\be
S^{(4)}  =  \frac{1}{2\a'} \int d^2 z (
 -\frac{1}{3} \dbar y^d y^c y^b \dif y^a R_{dcba}
 + \frac{1}{4} y^c y^d \de_d H_{cab} \dif y^a \dbar y^b )
\ee \label{sexpansion2} \es
and 

\bs\be
T^{(0)}  =  -\frac{1}{2\a'} \lambda_{ab} \P^a \P^b + \dif \P^a \Phi_a +
 \P^a \P^b \dif_b \Phi_a 
\ee\be
T^{(1)}  =  - \frac{1}{\a'}
\lambda_{ab} \P^a \dif y^b + \dif^2 y^a \Phi_a  
+  \dif y^a \Pi^b (\dif_{(b} \Phi_{a)} + \o_{[ba]}{}^c \Phi_c)
\ee\be
\eqalign{
T^{(2)}  = &  -\frac{1}{2\a'}
 \lambda_{ab} \dif y^a \dif y^b + \frac{1}{2\a'} \Pi^a \dif y^b y^c
(2 \lambda_b{}^e \hom_{aec} - \lambda_a{}^e H_{bec} )
\cr & -\frac{1}{4\a'} \Pi^a \P^b y^c y^d Z^4_{dcab} +
\frac{1}{2\a'} \Pi^d y^b y^c R_{dbc}{}^e \lambda_{ea} \Pi^a \cr}
\ee\be
\eqalign{
T^{(3)}  = & -\frac{1}{2\a'}  \P^a \dif y^b y^c y^d Z^5_{dcab}
 - \frac{1}{\a'} \dif y^a \dif y^b y^c \hop_{ca}{}^d \lambda_{db} \cr &
 +\frac{1}{6\a'} \dif y^d y^b y^c R_{dbc}{}^e \lambda_{ea} \P^a + 
 \frac{1}{2\a'} \P^d y^b y^c R_{dbc}{}^e \lambda_{ea} \dif y^a \cr }
\ee
\label{texpansion2}
\es

\noindent
where $\ho^{\pm}$ are the generalized connections (\ref{tordefb}) and
we have defined the following five objects
\bs\be
Z^1_{dcab} = \del_d H_{cab} + 2 \hom_{ace} \hop_{bd}{}^e + \frac{1}{2}
H_{ace} H_{bd}{}^e 
\ee\be
Z^2_{dcab} = \del_d H_{abc} - \hom_{ace} H_{db}{}^e - \frac{1}{2}
 H_{ace} H_{db}{}^e
\ee\be
Z^3_{dcab} = \del_d H_{bca} - \frac{1}{2} H_{bce} H_{da}{}^e + 
\hop_{bce} H_{da}{}^e 
\ee\be
\eqalign{ Z^4_{dcab} = & 
-\lambda_{af} H_{d}{}^{ef} \hom_{bec} + 2 \lambda^{ef} \hom_{afd} 
\hom_{bec} \cr &
+ \lambda_a{}^e (\del_d H_{bec} - \hom_{bcf} H_{de}{}^f -
\frac{1}{2} H_{bcf} H_{de}{}^f) \cr}
\ee\be
\eqalign{ Z^5_{dcab} =  \frac{1}{4} ( &
\lambda_a{}^f H_{dfe} H_{bc}{}^e - 2 \lambda_b{}^f \hom_{ace} H_{df}{}^e
 \cr & - 2 
\lambda^{ef} \hom_{afd} H_{bec} - 2 \lambda^{ef} H_{bfd} \hom_{aec}
 + \frac{1}{2} \lambda_{a}{}^e \del_d H_{bec} \cr & 
+\frac{1}{2}  \lambda_b{}^e (\del_d H_{aec} - 
\hom_{acf} H_{de}{}^f - \frac{1}{2}H_{acf} H_{de}{}^f ) ). \cr} 
\ee\label{defz} \es

\noindent
The background field expansions (\ref{sexpansion2}) and (\ref{texpansion2})
are the basis for the diagrammatic computations below.

\subsection{Momentum space and dimensional regularization}

The term $S^{(2)}$ of the action in (\ref{sexpansion2a}) 
contains the kinetic term
\be
S_{\rm kin} = \frac{1}{2\a'} \int d^2 z 
 \dif y_a \dbar y^a 
\label{skin}
\ee
which defines the coordinate-space propagators
%
\be
<y^a(z) y^b(w)> = \a' G^{ab} G(z,w) \equiv -\a' G^{ab} \log |z-w|^2 .
\label{prop1}
\ee
The remainder of the terms in the background field
expansion (\ref{sexpansion2}) define the vertices of the theory.

It is then straightforward to write down coordinate-space expressions
for the Feynman diagrams in Figure~1. 
In diagrams with two background fields, we can assume that the background
fields are independent of the coordinates, or, in momentum space,
that they have zero momentum. (In a Taylor expansion of the 
background fields around a fixed point on the 
world-sheet, all derivatives of these background fields have
higher conformal weight and can be ignored.) For diagrams with one
or zero background fields one has to be more careful and expand
everything up to the relevant order. 


In coordinate space, one then encounters various integrals not containing
background fields, which one could try to evaluate using identities such
as
\be \dif_{\bar{z}} \frac{1}{z-w} = \delta^{(2)}(z-w).
\label{fund}
\ee
However, one soon runs into problems with coordinate-space integrals
that give different answers when evaluated in different ways.
This ambiguity reflects the fact that one must regularize the 
(relatively convergent) integrals. Although 
a consistent all-order coordinate-space regularization [57--59]
is known for the general sigma model, and differential regularization 
\cite{freed} may also be applicable to this case,
we have chosen instead to use the more conventional method of
dimensional regularization in momentum space.

In dimensional regularization, we have to deal with several problems.
First, there are terms in the action coming from the antisymmetric
tensor field in (\ref{aux20a}), 
and these contain two-dimensional
antisymmetric tensors $\epsilon_{ij}$ when written in a world-sheet
covariant form. Furthermore, the stress tensor (\ref{aux20b}) is not
part of a covariant world-sheet tensor, because $\bar{T}$ contains
$\bar{\lambda}_{ab}$ which need not be equal to $\lambda_{ab}$. 
We have chosen the following version
of dimensional regularization to deal with these problems:
We keep all interactions in exactly two dimensions, 
but the kinetic terms 
(\ref{skin}), and hence the denominators of the Feynman diagrams, 
will be taken in $2-2\e$ dimensions.
With such a regularization, one has to be careful when going
beyond one
loop, in which case evanescent counterterms have to be taken into account
(see \cite{dewit} for a detailed discussion of two-loop
renormalization in a two-dimensional sigma model, including problems
with the $\e$ tensor in two dimensions).
Since our computation is limited to one loop, we will not discuss these
issues further here.

In parallel with our definition 
$d^2 z = {dxdy}/{\pi} $
we also
define $d^2k=dk_x dk_y /\pi$, and 
with this choice we see no explicit 
factors of $\pi$ in our calculation. Then $G(z,w)$ has the 
following representation
\be
G(z,w) = \int d^2 k \frac{e^{i\kb (z-w) + ik(\zb-\wb)}}{|k|^2}
\label{aux30}
\ee
and the momentum-space propagators read
\be
<y^a(k) y^b(l)>  =  \alpha' G^{ab} \delta^{(2)}(k+l) \frac{1}{|k|^2}
\label{propk}
\ee
where $k$ is shorthand notation for 
$k_z=(k_x+ i k_y)/2$, the $z$-component of $k$.

To work out the Feynman diagrams we write down the corresponding
expression in momentum space, continue only the denominators
to $d=2-2\e$ dimensions, and
include a factor 
$\Gamma(1-\e) (4\pi)^{-\e}(2\pi)^{2\e} \mu^{2\epsilon} $
 for each loop. The $\mu$ dependent factor is the usual 
dimensional regularization factor, necessary to maintain naive
dimensions in $d=2$, and the $\mu$ independent factors \cite{dewit}
remove many unwanted constants such as the Euler constant. The
$d$-dimensional measure $d^d k $ is the standard $d$-dimensional
measure divided by $\pi$. 
The relevant integrals 
\bea
 & & \Gamma(1-\e) (4\pi)^{-\e} (2\pi)^{2\e} \mu^{2\epsilon}
\int d^d k \frac{k^a \kb^b}{[|k|^2]^{\a} [|k-p|^2]^{\b} }
 =  \nonumber \\ & & \quad
 p^{a+1-\a-\b} \pb^{b+1-\a-\b} |p^2/\m^2|^{-\e}
 \frac{\Gamma[1-\e] \Gamma[1-\e-\a+b]}{\Gamma[\a] \Gamma[\b]
 \Gamma[2-\a-\b+b-\e]}  \nonumber   \\ & & \!\!\!\!
 \times \sum_{r=0}^{a}  \left(
 \begin{array}{c}  a   \\ r \end{array} \right)
 \frac{\Gamma[2-\a-\b+b+r-\e] \Gamma[\a+\b-1-r+\e] \Gamma[1-\e-\b+r]}{
 \Gamma[2-2\e-\a-\b+r+b]}
\label{dimreg}
\eea
are easily obtained by differentiating the known result for the
case $a=b=0$. 

In our one-loop calculation we encounter integrals like
$\int \frac{d^2 k }{|k|^2}$, which vanish in dimensional
regularization because of a cancellation between an ultraviolet 
and an infrared divergence.
In order to check in our calculation whether infrared and ultraviolet
divergences cancel separately, we have replaced the propagators
in each infrared divergent
diagram by
\be
\frac{1}{|k|^2} \rightarrow \frac{1}{|k|^2} + \frac{1}{\zeta} 
\delta^{(2)}(k) .
\label{irsub}
\ee
By taking $\zeta=\epsilon$, 
one can use this substitution to subtract out all
infrared divergences \cite{dewit}, but
we will keep $\zeta$ and $\epsilon$ 
as independent parameters. The 
infrared divergence of a diagram will then be given by the
coefficient of $-\frac{1}{\zeta} + \frac{1}{\epsilon}$, while the
ultraviolet divergence is given by the coefficient of
$\frac{1}{\epsilon}$.

We have worked out all our diagrams in momentum space using (\ref{dimreg})
and (\ref{irsub}),
and then afterwards reexpressed them in coordinate space using the
following translation table
\bs\be
-\frac{\pb^3}{3 {p}^3}  \leftrightarrow  
\frac{(\bar{z}-\bar{w})^2}{(z-w)^4} \label{transla} \ee\be
\frac{\pb^2}{2 {p}^2}  \leftrightarrow  
\frac{\bar{z}-\bar{w}}{(z-w)^3} \label{translb} \ee\be
-\frac{\pb}{{p}}  \leftrightarrow  
\frac{1}{(z-w)^2} \label{translc} \ee\be
-\frac{\pb}{\zeta p}+
\frac{\pb}{\epsilon {p}} + \frac{\pb}{{p}} (1-\log(|p|^2/\mu^2))
 \leftrightarrow  \frac{G(z,w)}{(z-w)^2} \label{transld} \ee\be
-\frac{\pb}{\zeta {p}}  \leftrightarrow  
\frac{G(0,0)}{(z-w)^2}.
\ee \label{transle}
\es
The first three relations
follow by Fourier transformation, and the next to last relation was
obtained
by comparing the
result for $<y(z)y(w)><\dif y(z) \dif y(w)>$ in coordinate space
and momentum space. On the left side of this relation,
one could have in principle chosen a different
finite term proportional to $p/\bar{p}$, but since all divergences
will cancel in our calculation, this would not make any
difference in the final results. Finally, the last relation was 
obtained from the momentum-space expression for $G(0,0)$ as it
follows from (\ref{aux30}).

\subsection{Diagrammatic results}

We now describe the results of the calculations for each of the diagrams
in Figure~1. Diagrams (1)-(13) are the one-loop contributions to the left hand
side of (\ref{aux20e}) involving no dilaton and two background fields,
diagrams (14)-(16) are the tree-diagrams involving the dilaton with 
two background fields, contributing to the left hand
side of (\ref{aux20e}), etc. Diagrams (1)-(20) and (24) are all
diagrams that contribute to $<T(z)T(w)>$,
and diagrams (21)-(23) are all diagrams that contribute to 
$<T(w)>$. Diagrams (1)-(23) all have the same order of $\alpha'$,
and (24) has one order of $\a'$ less. For completeness, we have included
in Fig.~1 
a set of other diagrams (25)-(37) that one could in
principle write down, but these do not contribute for a variety of reasons:
The 
tadpole diagrams (25)-(27) cancel algebraically against (28)-(30) in 
(\ref{aux20e}), while diagrams (31)-(33) 
vanish because they contain a contraction $H^{abc} \lambda_{bc}$.
Finally, 
 diagrams (34)-(37) contain integrals of the form $\int d^2 p 
p^a \bar{p}^b$ with $a\neq b$, which vanish identically as can
be seen by changing variables $p \rightarrow e^{i\phi} p$.

We now list the contributions of the individual diagrams, using the
notation $I_n$ for the $n$th diagram\footnote{$I_n$ contains
the contributions from all
diagrams that are topologically equivalent
to the $n$th diagram in Fig.~1, including, when
they are distinct, the mirror
diagrams obtained by the interchange $z\leftrightarrow w$.}
in Fig.~1. 
For diagrams (1)-(20) and
(24), $I_n$ is the contribution of this diagram to $<T(z)T(w)>$, while
for diagrams (21)-(23) $I_n$ is the contribution to $2<T(w)>/(z-w)^2$.
The results come out as follows:
\bs\bea
I_1 & = & \frb ({\lambda}^{ab} R_{bdca} -\half {\lambda}^{ab} \de_a H_{bcd} 
 \nonu & & \qquad
 - \hom_{cea} \hop_{d}{}^e{}_b {\lambda}^{ab}
 -\frac{1}{4} H_{cea} H_{d}{}^e{}_b {\lambda}^{ab} ) \\
I_2 & = & \fbb (\frac{1}{2} {\lambda}^{ae}
 {\lambda}^{bf} \hop_{cab} \hop_{def}) \nonu
 &  & +\frb (- {\lambda}^{ae}
 {\lambda}^{bf} \hom_{cab} \hop_{def}) \nonu
 &  & +\frr (\frac{1}{2} {\lambda}^{ae} {\lambda}^{bf} \hom_{cab} \hom_{def}) \\
I_3 & = & \fbb (-\frac{1}{2} {\lambda}^{ab} \hop_{caf} \hop_{db}{}^f ) \nonu
 &  & +\frb (2 {\lambda}^{ab} \hom_{caf} \hop_{db}{}^f ) \nonu
 &  & +\frrrg{-\frac{3}{2}}
  (- {\lambda}^{ab} \hom_{caf} \hom_{db}{}^f ) \\
I_4 & = & \frr ( \frac{1}{2} {\lambda}^{ab} {\lambda}_{c}{}^{f}
 H_{feb} \hom_{d}{}^e{}_{a}
 +  {\lambda}^{ab} {\lambda}^{ef} \hom_{cfb} \hom_{dea} \nonu
 & & \qquad \frac{1}{2} {\lambda}_{c}{}^{e} {\lambda}^{ab} \de_b H_{dea}
 +\frac{1}{2} {\lambda}_{c}{}^{e} {\lambda}^{ab} \hom_{dfa} H_{e}{}^f{}_b
 +\frac{1}{4} {\lambda}_{c}{}^{e} {\lambda}^{ab} H_{dfa} H_{e}{}^f{}_b
 \nonu
 & & \qquad 
 - {\lambda}^{ab} {\lambda}_{c}{}^{e} R_{bdea} ) \\
I_5 & = &
 \frr ( \frac{1}{4} {\lambda}_{c}{}^{e} {\lambda}_{d}{}^{f} 
 H_{eab} H_{f}{}^{ab}
 + {\lambda}_{c}{}^{e} {\lambda}^{af} H_{eab} \hom_{df}{}^{b}
 +  {\lambda}^{be} {\lambda}^{af} \hom_{cea} \hom_{dbf}) \nonu
& & \!\!\!\!\!\! +
 \frrg ( -\frac{1}{4} {\lambda}_{c}{}^{e} {\lambda}_{d}{}^{f}
  H_{eab} H_{f}{}^{ab}
 - {\lambda}_{c}{}^{e} {\lambda}^{bf} H_{eba} \hom_{df}{}^{a}
 -  {\lambda}^{ef} \hom_{cea} \hom_{df}{}^{a}) \\
I_6 & = & \frb({2} {\lambda}^{ab} {\lambda}^{ef} \hom_{ceb} \hop_{dfa}
 -2 {\lambda}^{be} \hom_{cef} \hop_{db}{}^{f}) \nonu
& & + \frr ( - {\lambda}^{ab} {\lambda}_{c}{}^{e} H_{ebf} \hom_{da}{}^{f}
 -2 {\lambda}^{ab} {\lambda}^{ef} \hom_{ceb} \hom_{dfa} ) \nonu
& & + \frrrg{-1} (2 {\lambda}^{be} \hom_{cef} \hom_{db}{}^{f}
 + {\lambda}^{ab} {\lambda}_{c}{}^{e} H_{ebf} \hom_{da}{}^{f} ) \\
I_7 & = & \frrd (
 \frac{1}{4} {\lambda}_{d}{}^{b} {\lambda}_{c}{}^{f} H_{bea} 
  H_{f}{}^e{}_{a} -
 \frac{1}{2} {\lambda}_{d}{}^{f} H_{fea} \hom_{c}{}^{ea} +
  {\lambda}_{d}{}^{b} {\lambda}^{ef} \hom_{cfa} H_{be}{}^{a}  \nonu
& & \qquad 
 -\half {\lambda}_{d}{}^{e} \de_a H_{ce}{}^{a}
 -\frac{1}{2} {\lambda}_{d}{}^{e} \hom_{cfa} H_{e}{}^{fa}
 -\frac{1}{4} {\lambda}_{d}{}^{e} H_{cfa} H_{e}{}^{fa}  \nonu
& & \qquad +\frac{1}{3} {\lambda}_{d}{}^{b} {\lambda}_{c}{}^{e} R_{abe}{}^{a}
 + {\lambda}_{d}{}^{e} R_{ace}{}^{a} ) \\
I_8 & = & \frrd(
 - \hom_{cea} H_{b}{}^e{}_{f} {\lambda}^{af} {\lambda}_{d}{}^{b}
 + {\lambda}_{d}{}^{f} \hom_{cea} H_{f}{}^{ea} )  \\
I_9 & = & \frrrg{-1}(-\frac{1}{4}
 {\lambda}_{c}{}^{a} {\lambda}_{d}{}^{b} H_{aef} 
 H_{b}{}^{ef} ) \nonu
& & + \frrd( \frac{1}{4} {\lambda}_{c}{}^{a} {\lambda}_{d}{}^{b} 
  H_{aef} H_{b}{}^{ef} ) \\
I_{10} & = & \frrd (-\frac{1}{3} {\lambda}_{c}{}^{a} {\lambda}_{d}{}^{e}
   R_{bea}{}^{b} ) \\
I_{11} & = & \frr({\lambda}_{c}{}^{f} {\lambda}^{be} R_{efdb}
 -\half {\lambda}_{c}{}^{f} {\lambda}^{be} \de_e H_{dfb}  \nonu
& & \qquad
 -\frac{1}{2} {\lambda}_{c}{}^{f} {\lambda}^{be} \hom_{dab} 
  H_{f}{}^a{}_{e}
 -\frac{1}{4} {\lambda}_{c}{}^{f} {\lambda}^{be} H_{dab} 
  H_{f}{}^a{}_{e} ) \nonu
& & + \frb (
 -2 {\lambda}_{c}{}^{f} {\lambda}^{be} R_{efdb}
 + {\lambda}_{c}{}^{f} {\lambda}^{be} \de_e H_{dbf}  \nonu
& & \qquad
 +\frac{1}{2} {\lambda}_{c}{}^{f} {\lambda}^{be} H_{dab} 
  H_{f}{}^a{}_{e}
 - {\lambda}_{c}{}^{f} {\lambda}^{be} \hop_{dab} H_{f}{}^a{}_{e} ) \\
I_{12} & = & \frb ( {\lambda}_{c}{}^{b}
 {\lambda}^{ah} H_{bfa} \hop_{d}{}^f{}_{h}) \nonu
 & & + \frrrg{-\frac{3}{2}} (- {\lambda}_{c}{}^{b}
 {\lambda}^{ah} H_{baf} \hom_{dh}{}^{f}) \nonu
 & & + \frrd ( {\lambda}_{c}{}^{b}
 {\lambda}^{ah} H_{baf} \hom_{dh}{}^{f}) \\
I_{13} & = &
\frrrg{-1}(\half {\lambda}_{c}{}^{b} {\lambda}_{d}{}^{f} H_{bae} H_{f}{}^{ae}
 + {\lambda}_{c}{}^{b} {\lambda}^{af} H_{bae} \hom_{df}{}^{e} ) \nonu
 & & + \frrd(-\half {\lambda}_{c}{}^{b} {\lambda}_{d}{}^{f} H_{bae} 
  H_{f}{}^{ae}
 - {\lambda}_{c}{}^{b} {\lambda}^{af} H_{bae} \hom_{df}{}^{e} ) \\
I_{14} & = &
 \frac{1}{(z-w)^2} (4 \P^c {\lambda}_{c}{}^{a} \pa( \Phi_a)
 +2  \Pi^c \Pi^d 
 {\lambda}_{c}{}^{a} \del_{[a} \Phi_{d]}  ) \\
I_{15} & = &
 \frb (  {\lambda}_{c}{}^{a}  \Phi_b \hop_{d}{}^{ab} ) \nonu
 & & +\frr (-2  {\lambda}_{c}{}^{a} \Phi_b \hom_{d}{}^{ab} ) \\
I_{16} & = &
 +\frr (   {\lambda}_{c}{}^{a} \Phi_b \hom_{d}{}^{ab}
   +  {\lambda}_{d}{}^{a} \Phi_b \hom_{c}{}^{ab}  )  \\
I_{17} & = & \fbb(
  {\lambda}^{fb} \hop_{cef} \hop_{d}{}^e{}_{b}
 - {\lambda}^{eb} {\lambda}^{af} \hop_{dab} \hop_{cfe} ) \nonu
& & + \frb(
 - {\lambda}^{fb} \hom_{cef} \hop_{d}{}^e{}_{b}
 + {\lambda}^{eb} {\lambda}^{af} \hop_{dab} \hom_{cfe} \nonu
& & \qquad
 + {\lambda}^{fb} \hop_{cef} \hop_{d}{}^e{}_{b}
 - {\lambda}^{eb} {\lambda}^{af} \hop_{dab} \hop_{cfe} ) \nonu
& & + \frr( - {\lambda}^{fb} \hom_{cef} \hop_{d}{}^e{}_{b}
 + {\lambda}^{eb} {\lambda}^{af} \hop_{dab} \hom_{cfe} ) \\
I_{18} & = & \frb(
 2 {\lambda}^{fb} \hom_{cef} \hop_{d}{}^e{}_{b}
 -2 {\lambda}^{eb} {\lambda}^{af} \hop_{dab} \hom_{cfe} ) \nonu
& & + \frr (
 2 {\lambda}^{fb} \hom_{cef} \hop_{d}{}^e{}_{b}
 -2 {\lambda}^{eb} {\lambda}^{af} \hop_{dab} \hom_{cfe} ) \\
I_{19} & = & \frr (
 2{\lambda}_{c}{}^{e} \hop_{d}{}^b{}_{e} \Phi_b 
  + 2{\lambda}^{be} \hop_{dce} \Phi_b )
 \nonu & & 
 + \frac{1}{(z-w)^2} (
 +2 {\lambda}_{c}{}^{b} \pa \P^c \Phi_b
 -2 {\lambda}_{c}{}^{b} \P^c \pa  \Phi_b) \nonu
& & + \frb( 2
 {\lambda}_{c}{}^{e} \hop_{d}{}^b{}_{e} \Phi_b ) \nonu
& & + \frac{(\zb-\wb)}{(z-w)^3} (
 -2 {\lambda}_{c}{}^{b} \P^c \difb  \Phi_b) \\
I_{20} & = &
 \frac{1}{(z-w)^4} (\frac{1}{2} \lambda_a{}^b
 \lambda_b{}^a ) \nonu
 & & +
 \fbb(
 -\frac{1}{2} {\lambda}^{fb} \hop_{cef} \hop_{d}{}^e{}_{b}
 +\frac{1}{2} {\lambda}^{eb} {\lambda}^{af} \hop_{dab} \hop_{cfe} ) \nonu
& & + \frb(
 - {\lambda}^{fb} \hop_{cef} \hop_{d}{}^e{}_{b}
 + {\lambda}^{eb} {\lambda}^{af} \hop_{dab} \hop_{cfe} ) \nonu
& & + \frr( -\frac{1}{2} {\lambda}^{fb} \hop_{cef} \hop_{d}{}^e{}_{b}
 +\frac{1}{2} {\lambda}^{eb} {\lambda}^{af} \hop_{dab} \hop_{cfe} ) \\
I_{21} & = & \frrd ( - {\lambda}^{ab} \hom_{cae} \hom_{db}{}^{e} ) \\
I_{22} & = & \frrd (  {\lambda}_{c}{}^{e} \hom_{dab} H_{e}{}^{ab}
 +2 {\lambda}^{be} \hom_{dab} \hom_{c}{}^a{}_{e} ) \\
I_{23} & = & \frrd ({\lambda}_{c}{}^{e} R_{bde}{}^{b}
 -\frac{1}{2} {\lambda}_{c}{}^{f} H_{fea} \hom_{d}{}^{ea}
 - {\lambda}^{ef} \hom_{cfa} \hom_{de}{}^{a} \nonu
& & \qquad -\half {\lambda}_{c}{}^{e}
 \de_a H_{de}{}^{a} -\frac{1}{2} {\lambda}_{c}{}^{e} \hom_{dfa} H_{e}{}^{fa}
 -\frac{1}{4} {\lambda}_{c}{}^{e} H_{dfa} H_{e}{}^{fa} ) \\
  I_{24}  & =  & \frr ( -\frac{1}{\a'} {\lambda}_{c}{}^{a} {\lambda}_{da} ).
\eea \label{alldiag} \es

The background fields in these results are evaluated at the point
$(w,\bar{w})$.

\subsection{Derivation of the unified system}

With the diagrammatic results (\ref{alldiag}), we can now obtain
the unified Einstein-Virasoro master equation at the one-loop level.
If we insert
the results from the diagrams into (\ref{aux20e}), and we include also
the classical expectation value of $T$
\be
<T(w)>_{\rm cl} = T^{(0)}(w) \equiv 
-\frac{1}{2\a'} \l_{ab} \Pi^a \Pi^b + \dif(\Pi^a \Phi_a),
\ee
we obtain an equation of the form
\bs
\bea
\!\!\!\! \oa & \!\!\!\!=& \!\!\!\!
<T(z)T(w)> - \frac{c/2}{(z-w)^4} - 2 \frac{<T(w)>}{(z-w)^2} 
\\ \!\!\!\! & \!\!\!\!=  & \!\!\!\!
 A \frac{1}{(z-w)^4} + (\Pi^c \bar{\Pi}^d B_{cd}) 
\frac{(\zb-\wb)}{(z-w)^3} + (\Pi^c \Pi^d C_{cd}
+ \dif(\Pi^c E_c)
)\frac{1}{(z-w)^2} \label{fineqb}
\eea
\label{fineq}
\es
in which, quite remarkably, all terms proportional to $(\zb-\wb)^2/(z-w)^4$,
$G(z,w)$ and $G(0,0)$ cancel in a highly non-trivial fashion. The fact
that the infrared and ultraviolet divergences cancel separately,
which one would
naively expect for a conserved current like $T$,
is an excellent check on the diagrammatic results (\ref{alldiag}). 
After a certain amount of 
algebra, we find the following expressions for $A$, $B_{cd}$, $C_{cd}$ and
$E_c$
\bs\bea
A & = & \frac{1}{2} \lambda_a{}^b \lambda_b{}^a - \frac{c}{2} =\oa 
\label{tensa}
\\{}
B_{cd} & = & \lambda^{ab}( \hat{R}^+_{dabc} - 
 2 \lambda_c{}^f  \hat{R}^+_{dabf}) - 
 2 \lambda_{c}{}^b \delhp_d \Phi_b = \oa 
\label{tensb}
\\{}
C_{cd} & = & \frac{1}{\a'}(\lambda_{cd} - \lambda_c{}^a \lambda_{ad} ) +
\frac{1}{2} \lambda^{ae} \lambda^{bf} H_{cab} H_{def}  - 
\frac{1}{2} \lambda^{ef} H_{cbe} H_d{}^b{}_f 
\nonu & & 
+ \lambda_c{}^a \del_{[a} \Phi_{d]} 
+ \lambda_d{}^a \del_{[a} \Phi_{c]} = \oa \label{tensc}
\\{}
E_c & = & 2(\lambda_c{}^a \Phi_a - \Phi_c) = \oa
\label{newdilrel}
\eea\label{tens} \es
Each of these tensors must vanish separately at the indicated order
because they multiply independent structures in (\ref{fineqb}).
These covariant results show that all spin connections, and hence
anomalies, cancel in the final result, as expected
in the absence of chiral fermions. 

\vs .4cm
\noindent \underline{Eigenvalue relation for the dilaton vector}
\vs .3cm 

Eq. (\ref{newdilrel}) is recognized (with $\lambda_a{}^b = 2L_a{}^b$)
as the eigenvalue relation (\ref{evn}) of the dilaton eigenvector.
Using (\ref{aux20d}), a
particular solution of the eigenvalue relation is the dilaton solution
$\Phi_a = \Phi_a(\Phi) =2L_a{}^b \del_b \Phi$ in (\ref{eig}).

\vs .4cm
\noindent \underline{Generalized Einstein equations}
\vs .3cm 

From (\ref{tensb}) we derive that
$B_{cd} - 2 \lambda_c{}^g B_{gd} = \oa$, which yields
\be
 \lambda^{ab} \hat{R}^+_{dabc} + 2 \lambda_c{}^b \delhp_d \Phi_b = \oa
\label{jp1}
\ee
as an equivalent form of (\ref{tensb}).
Using (\ref{newdilrel}) and (\ref{aux20d}), we find that 
(\ref{jp1}) is also
equivalent to 
\be
 \lambda^{ab} \hat{R}^+_{dabc} + 2 \delhp_d \Phi_c = \oa
\label{jpq1}
\ee
which is recognized
as the generalized Einstein equation (\ref{eva}).

%
%

\vs .4cm
\noindent \underline{Central charge}
\vs .3cm 

In (\ref{tensa}) we recognize the classical limit of the central
charge in (\ref{evd}),
which was in fact determined by using the Bianchi identities 
and the fact that the central charges must be constant (see
Section~5.7). It should also be possible to verify
the order $\a'$ terms of (\ref{evd}) 
in a two-loop calculation, but we have
not attempted to do this because the renormalization of the sigma model
becomes quite subtle at two loops. 

We can however make a few simple remarks in this direction.
In addition to the two-loop contributions,
there are also one-loop diagrams with one dilaton vector insertion
and tree diagrams with two dilaton vector insertions.
These diagrams
are easily evaluated and we find that they contribute to 
the coefficient $A$ in
the form
\bs
\bea A  & = & 
 \frac{1}{2} \lambda_a{}^b \lambda_b{}^a - \frac{c}{2} \\{}
& & + 6 \a' ( \Phi_a \Phi^a - \delhp_a \Phi^a) \\{}
& & + \mbox{{\rm two-loop contributions}} \\{}
& =  &  \oaa.
\eea \es
Looking at the alternate forms of the central
charge in (\ref{varc}), we see only one form consistent
with these terms in $A$, namely (\ref{varcf}). Thus, we expect that
a complete two-loop computation of the central charge will give the
form (\ref{varcf}) directly.

\vs .4cm
\noindent \underline{Generalized VME}
\vs .3cm 

This leaves equation (\ref{tensc}), which, except in the classical limit,
 does not agree in form
with the generalized VME in (\ref{evc}), and, in particular,
(\ref{tensc}) does not
agree with the VME when the sigma model
is the WZW action. The VME is derived using
OPEs, which is a particular way of
regularizing the WZW model. Since different regularization schemes can
differ only by finite local counterterms, we are led to look
for a finite local counterterm which restores the form of the VME.
Such a counterterm is easily found, namely if we define
\be T' = T - \frac{1}{4} \Pi^c \Pi^d \lambda^{ef} H_{dbf} H_{c}{}^b{}_e
\ee
and use $T'$ rather than $T$ in (\ref{aux20e}), we find a few extra
contributions to $C_{cd}$. In particular, we
 have to include the counterterm in
the $T^{(0)}(w)$ contribution to $<T(w)>$, and in addition there
are tree diagrams with an insertion of the counterterm contributing to
$<T(z) T(w)>$. With these extra terms, the coefficient  $C_{cd}$ becomes
\bea
C_{cd} & = & \frac{1}{\a'}(\lambda_{cd} - \lambda_c{}^a \lambda_{ad} ) +
\frac{1}{2} \lambda^{ae} \lambda^{bf} H_{cab} H_{def} \nonu
& &  - \frac{1}{2} \lambda^a{}_{(c} H_{d)be} H_{a}{}^b{}_f \lambda^{ef}
+ \lambda_c{}^a \del_{[a} \Phi_{d]} 
+ \lambda_d{}^a \del_{[a} \Phi_{c]} \nonu
& = & \oa
\eea
which is recognized as
the generalized VME (\ref{evc}). This concludes the derivation
of the unified Einstein-Virasoro master equation at the one-loop
level.

\section{Conclusions}

We have studied the general Virasoro construction
\bs\be 
T=-\frac{L_{ij}}{\a'} \partial x^j \partial x^j + {\cal O}(\a'^0)
\ee\be
i,j=1,\ldots,\dim(M)
\ee\es
at one loop
in the operator algebra of the general non-linear sigma model,
where $L$ is a spin-two spacetime tensor field called the inverse
inertia tensor. The construction is summarized by a unified 
Einstein-Virasoro master equation which describes the covariant coupling of 
$L$ to the spacetime fields $G$, $B$ and $\Phi_a$, where $G$ and $B$
are the metric and antisymmetric tensor of the sigma model and
$\Phi_a$ is the dilaton vector, which generalizes the derivative
$\del_a \Phi$ of the dilaton $\Phi$.
As special cases, the unified system contains the Virasoro master equation
of the general affine-Virasoro construction 
and the conventional Einstein equations of the canonical sigma model
stress tensors.  More generally, the unified system describes a space of
 conformal field theories which is presumably much larger than the sum of
these two special cases.
    
We have also discussed a number of algebraic and geometric properties
 of the system, noting in particular the relation
of the system to an unsolved problem in 
the theory of $G$-structures on manifolds 
with torsion.

In addition to questions posed in the text,
we list here a number of other important directions.

\noindent
1. New solutions. 
It is important to find new solutions of the unified system, beyond 
the canonical stress tensors of the sigma model 
and the general affine-Virasoro construction. 
Here the question of $G$-structures 
(see Section~5.5) and solution classes (see Section~5.8)
will be important.

\noindent
2. Duality. 
The unified system contains the coset constructions in two distinct 
ways, that is, both as $G_{ab}=k \eta_{ab}$, $L^{ab}=L^{ab}_{g/h}$
 in the general affine-Virasoro
construction and among the canonical stress tensors of
the sigma model with the sigma model metric that corresponds to the
coset construction. 
This is an indicator of new duality transformations in
the system, possibly exchanging $L$ and $G$, 
which may go beyond the coset constructions. 
In this connection, we 
remind the reader that the VME has been identified
\cite{gme} as an Einstein-Maxwell system with torsion on the group manifold, 
where the inverse inertia tensor is the inverse metric on
tangent space. Following this hint, it may be possible to cast the 
unified system on group manifolds as two coupled Einstein
systems, with exact covariant constancy of both $G$ and $L$.

\noindent
3. Spacetime action and/or $C$-function. 
These have not yet been found for the unified system,
but we remark that they are known for the
special cases unified here: The spacetime action
\cite{sig9,ghor} is known for the conventional Einstein equations of the
sigma model, and, for this case,
 the $C$-function is known \cite{sig13} for constant dilaton.
Moreover, an exact $C$-function is known  \cite{cf}
for the special
case of the unimproved VME.

\noindent
4. World-sheet actions. We have studied here only the Virasoro 
operators constructible in the operator algebra of the general sigma model,
but we have not yet worked out the world-sheet actions of the 
corresponding new conformal field theories, whose beta functions
should be the unified Einstein-Virasoro master equation. 
This is a familiar situation in
the general affine-Virasoro construction, whose Virasoro operators 
are constructed in the operator algebra of the WZW model, while
the world-sheet actions of the corresponding new conformal field theories
include spin-one \cite{br} gauged WZW models for the coset 
constructions and spin-two \cite{hy,ts3,bch}
 gauged WZW models for the generic construction. 

As a consequence of 
this development in the general affine-Virasoro construction, 
more or less standard Hamiltonian
methods now exist for the systematic construction of the new world-sheet
actions from the new stress 
tensors, and we know for example that $K$-conjugation
covariance is the source of the spin-two gauge invariance in the generic case.
At least at one loop, a large subset of Class IIb solutions of the
unified system exhibit $K$-conjugation covariance (see Section~5.9),
so we may reasonably expect that the world-sheet
actions for generic constructions in this subset are
spin-two gauged sigma models. For solutions with no
$K$-conjugation covariance, the possibility remains open
that these constructions
 are dual descriptions of other conformal sigma models.

\noindent
5. Superconformal extensions. The method of Ref. \cite{sig12}
has been extended [64--66] to the
canonical stress tensors of the supersymmetric
sigma model. The path is therefore open to study  general 
superconformal constructions in the operator algebra of the general
sigma model with fermions.
Such superconformal extensions should then include and 
generalize the known $N=1$ and $N=2$ superconformal master
equations \cite{sme} of the general affine-Virasoro construction. 

In this connection, we should mention that that the Virasoro master
equation is the true master equation, because it includes as a small 
subspace all the solutions of the superconformal master
equations. It is reasonable to expect therefore that, in the same way, 
the unified system of this paper will include the superconformal
extensions.

\noindent
6. Tree-level dilaton vectors. We have assumed 
in our computation
 that the dilaton
vector, like the dilaton, satisfies $\Phi_a={\cal O}(\a'^0)$, so
that its contribution begins at one loop. This assumption can be relaxed
to include dilaton vectors which begin at ${\cal O}(\a'^{-1})$ and
hence contribute at tree level. As noted in Section~5.2, these dilaton
vectors correspond to improvement vectors $D^a={\cal O}(k^0)$ in the
improved VME. We have not done this computation, but we expect that
the unified system remains the same in this case, except that one will
now see explicitly the conjectured ${\cal O}(\a')$ terms in the 
eigenvalue relation (\ref{dileig}) for the dilaton vector.

\addcontentsline{toc}{section}{Acknowledgements}
\section*{Acknowledgements} For many helpful discussions, we would
like to thank our colleagues:
L. Alvarez-Gaum\'e, I. Bakas,
K. Clubok, S. Deser,  A. Giveon, 
R. Khuri, E. Kiritsis, 
P. van Nieuwenhuizen,
N. Obers, A. Sen, K. Skenderis,
A. Tseytlin and B. de Wit.
We also thank CERN for its
hospitality and support during the course of this work.

The work of MBH was supported in part by the Director, Office of
Energy Research, Office of High Energy and Nuclear Physics, Division of
High Energy Physics of the U.S. Department of Energy under Contract
DE-AC03-76SF00098 and in part by the National Science Foundation under
grant PHY90-21139. The work of JdB was supported in part by
the National Science Foundation under grant PHY93-09888.

\vskip 1.0cm
\setcounter{equation}{0}
\def\theequation{A.\arabic{equation}}
 \boldmath
\addcontentsline{toc}{section}{Appendix A:
Chiral currents in the general sigma model}
\centerline{\bf Appendix A: Chiral currents in the general sigma model}
\unboldmath
\vskip 0.5cm
In this appendix, we construct a set of non-local chiral currents
in the general classical non-linear sigma model, which reduce to the
currents of affine $g \times g$ in the special case of the WZW model.

The Minkowski space currents $J^{\pm}$ of Section~3 satisfy the general
current algebra (\ref{alg}) and the classical equations of motion
\bs\be
\dif_{\mp} J^{\pm}_a = (M_{\pm})_a{}^b J_b^{\pm} 
\label{aux22a} \ee\be
(M_{\pm})_a{}^b = J_c^{\mp} (\hat{\omega}^{\pm c})_a{}^b =
\dif_{\mp} x^i (\ho^{\pm}_i)_a{}^b. \label{aux22b}
\ee \label{aux22} \es
where $a,b=1,\ldots,\dim(M)$. These relations are the Minkowski-space
analogues of the Euclidean equations of motion in (\ref{eqmot}).
In what follows, we will also need the antiordered exponentials
 $(R_{\pm})_a{}^b$,
which satisfy
\bs\be
\dif_{\mp} R_{\pm} + R_{\pm} M_{\pm}=0 \label{aux23a}
\ee\be
R_{\pm}(\xi) = f_{\pm}(\xi^{\pm}) T_{\mp}^{\ast} 
e^{-\int^{\xi^{\mp}} d\xi'{}^{\mp} M_{\pm}(\xi^{\pm},\xi'{}^{\mp}) }
\label{aux23b} \ee\be
(R_{\pm})_a{}^c (R_{\pm})_b{}^d G_{cd} = G_{ab} \label{aux23c}
\ee  \label{aux23} \es
where $T^{\ast}_{\mp}$ is antiordering 
in $\xi^{\mp}$. The matrices
$f_{\pm}$ are taken to be
pseudoorthogonal, so that $R_{\pm}$ also satisfies
(\ref{aux23c}).

The chiral currents $\J^{\pm}$ of the general sigma model 
are then constructed as
\bs\be
\J_a^{\pm} = \pm (R_{\pm})_a{}^b J_b^{\pm} 
\label{aux24a} \ee\be
\dif_{\mp} \J_a^{\pm} = 0 
\label{aux24b} \ee \label{aux24} \es
where the chirality condition (\ref{aux24b}) follows directly from
(\ref{aux22}) and (\ref{aux23}). We have checked that these chiral
currents are Noether currents corresponding to the non-local
symmetries of the sigma-model action (\ref{auxn16}),
\be 
\delta_{\pm} x^i = 4\pi \a' \e^a_{\pm} (\xi^{\pm}) (R_{\pm})_a{}^i,
\qquad \delta S=0
\ee
where $(R_{\pm})_a{}^i = (R_{\pm})_a{}^b e_b{}^i$ and 
$\epsilon_{\pm}(\xi_{\pm})$ are the parameters of the 
symmetry
transformations.

As a check on our construction, we consider the general chiral 
currents (\ref{aux24}) in the special case of flat generalized
connections,
\be \hat{R}^{\pm}_{abcd}=0 \ee
which includes the case of WZW. When the
generalized connections are flat, (\ref{aux23}) may be solved in
the form 
\be \dif_i (R_{\pm})_a{}^b + (R_{\pm})_a{}^c 
(\ho^{\pm}_i)_c{}^a = 0
\label{aux25} \ee
so that $R_{\pm}(x)$ are matrix-valued
functions of the coordinates $x^i(\xi)$ of the
sigma model. Then we obtain the equal-time brackets
\bs\be
[J_a^{+}(\s),R_{\pm}(\s ')_b{}^c]=
[J_a^{-}(\s),R_{\pm}(\s ')_b{}^c]=4\pi i \a' \delta(\s-\s ')
(R_{\pm})_b{}^d (\ho^{\pm}_a)_d{}^c
\ee\be
[R_+(\s),R_{\pm}(\s ')]=
[R_-(\s),R_{\pm}(\s ')]=0 \ee\es
from (\ref{aux25}) and the fundamental brackets (\ref{fundalg}).

Using these brackets and the current algebra (\ref{alg}), the algebra
of the chiral currents is computed as
\bs\be
[\J^{\pm}_a(\s),\J_b^{\pm}(\s ')] = 
4 \pi i \a' \delta(\s - \s ') 
{\cal F}_{ab}^{\pm}{}^c \J_c^{\pm}(\sigma) \pm
8 \pi i \a' G_{ab} \dif_{\s} \delta(\s-\s') 
\label{aux26a} \ee\be
[\J^{+}_a(\s),\J_b^{-}(\s ')] = 0 
\label{aux26b} \ee
\vskip -.15in
\be
{\cal F}_{abc}^{\pm} = (R_{\pm})_a{}^d (R_{\pm})_b{}^e
 (R_{\pm})_c{}^f H_{def}
\label{aux26c} \ee \label{aux26} \es
This algebra can be identified as the bracket form of affine $g \times g$
because the structure constants ${\cal F}_{\pm}$ in (\ref{aux26c})
are indeed constants
\be
\dif_i {\cal F}_{abc}^{\pm} = 
 (R_{\pm})_a{}^d (R_{\pm})_b{}^e
 (R_{\pm})_c{}^f \delh^{\pm}_i H_{def}=0
\ee
when the generalized connections are flat (see (\ref{cyclic})).

This result shows that all sigma model actions with $\hat{R}^{\pm}_{abcd}=0$
(that is, all sigma models on parallelizable 
manifolds) are equivalent to
the WZW action.
Although this statement is apparently well known, we have been unable to
find any proof in the literature.
In the standard WZW frame (\ref{auxn17}), we may solve
the differential equations
 (\ref{aux25})
explicitly with the results
\bs\be
(R_+)_a{}^b = \delta_a{}^b, \qquad (R_-)_a{}^b = (\Omega^{-1})_a{}^b
\ee\be
\J_a^+ = J_a^+, \qquad \J^-_a= -(\Omega^{-1})_a{}^b J_b^-
\ee\be
{\cal F}^{\pm}_{ab}{}^c = \frac{1}{\sqrt{\a'}} f_{ab}{}^c
\ee\es
where $\Omega\in{\rm Aut}(g)$ is the adjoint action of $g$
(see eqs (\ref{aux3}) and (\ref{aux4})). Comparing with
the WZW data (\ref{auxn1}), we see that
$\J^+$ and $\J^-$ are the Minkowski-space versions of the standard
holomorphic and anti-holomorphic WZW currents $J$ and $\bar{J}$.

It is an interesting open problem to find the algebra of the chiral
currents (\ref{aux24}) beyond the case of flat connections.

\end{document}